\documentclass[preprint]{aastex}

\usepackage{graphicx,epsfig,fancyhdr,rotating,amsmath,natbib}

\def\ie{{ i.e. }}
\newcommand{\degree}{\ensuremath{^\circ}}
\usepackage{txfonts}

\shorttitle{Accelerating waves in polar coronal holes}
\shortauthors{Gupta et al.}
\begin{document}
 

\title{Accelerating waves in polar coronal holes as seen 
by EIS and SUMER}


\author{G. R. Gupta\altaffilmark{1,2}, D. Banerjee\altaffilmark{1}, L. Teriaca\altaffilmark{3}, S. Imada\altaffilmark{4},
 S. Solanki\altaffilmark{3,5}}
\altaffiltext{1}{Indian Institute of Astrophysics, Koramangala, Bangalore 560034, India}
\altaffiltext{2}{Joint Astronomy Programme, Indian Institute of science, Bangalore 560012, India}
\email{girjesh@iiap.res.in}
\altaffiltext{3}{Max-Planck-Institut f\"{u}r Sonnensystemforschung (MPS), 37191  
     Katlenburg-Lindau, Germany}
\altaffiltext{4}{Institute of Space and Astronautical Science, Japan Aerospace Exploration Agency, Kanagawa, Japan}
\altaffiltext{5}{School of Space Research, Kyung Hee University, Yongin, Gyeonggi 446-701, Korea}

\begin{abstract}
We present EIS/Hinode \& SUMER/SoHO observations of propagating disturbances 
detected in coronal lines in inter-plume and plume regions of a polar coronal 
hole. The observation was carried out on $13^{th}$ November 2007 as
JOP196/HOP045 programme.
The SUMER spectroscopic observation gives the information about the fluctuation
in radiance and on both resolved (Doppler shift) and unresolved (Doppler width) line-of-sight
velocities whereas EIS $40\arcsec$ wide slot images detect 
fluctuations only in radiance but maximizes the probability of overlapping field of view between 
the two instruments. 
From distance-time radiance maps, we detect the presence of propagating waves 
in a polar inter-plume region with a period of 15~min to 20~min and a propagation 
speed increasing from ($130~\pm~14$)~km~s$^{-1}$\ just above the limb, to
($330~\pm~140$)~km~s$^{-1}$\ around $160\arcsec$\ above the limb.
These waves can be traced to originate from a bright region of the on-disk 
part of the coronal hole where the propagation speed is in the range of 
($25~\pm~1.3$)~km~s$^{-1}$ to ($38~\pm~4.5$)~km~s$^{-1}$, with the same 
periodicity. 
These on-disk bright regions can be visualized as the base of 
the coronal funnels. The adjacent plume region also shows the presence of 
propagating disturbance with the same range of periodicity but with propagation 
speeds in the range of ($135~\pm~18$)~km~s$^{-1}$ to ($165~\pm~43$)~km~s$^{-1}$\ only. 
A comparison between the distance-time radiance map of both regions, indicate 
that the waves within the plumes are not observable (may be getting dissipated) 
far off-limb whereas this is not the case in the inter-plume region. 
A correlation analysis was also performed to find out the time delay between 
the oscillations at several heights in the off-limb region, finding results
consistent with those from the analysis of the distance-time maps.
To our knowledge, this result provides first spectroscopic evidence of 
acceleration of propagating disturbances in the polar region close to the Sun 
(within 1.2~R/R$_{\odot}$), which provides clues to the understanding of the 
origin of these waves. 
We suggest that the waves are likely either Alfv\'{e}nic or fast magnetoacoustic in the 
inter-plume and slow magnetoacoustic in plume regions.
This may lead to the conclusion that inter-plumes are preferred channel for 
the acceleration of the fast solar wind.\\
 
\end{abstract}


\section{Introduction}
Coronal holes are regions of cool and low density plasma that, as such, are
`dark' at coronal temperatures \citep{1972ApJ...176..511M}.
During solar minima, coronal holes are generally confined
to the Sun's polar regions, while at solar maxima they can also be found
at lower latitudes, usually associated with remnant active regions, as
so-called `equatorial' coronal holes. The predominantly unipolar
magnetic field from coronal hole regions is thought to give rise
to the fast solar wind \citep[e.g.,][]{1973SoPh...29..505K}. 
During solar minimum, Ulysses observations clearly show that
the solar wind exhibits two modes of outflow: the fast wind, associated with 
polar coronal holes, with outflow speeds of $\approx800$~km~s$^{-1}$\ and the 
slow  wind with outflow speeds of $\approx400$~km~s$^{-1}$\ associated with 
equatorial regions \citep{1997GeoRL..24.2885W,2000JGR...10510419M}. 
However, during solar maxima, low latitude coronal holes
 also show faster than average solar wind speed upto 
 $\approx600$~km~s$^{-1}$ \citep{2003JGRA..108.1144Z}. 
Extreme-ultraviolet images of polar coronal 
holes reveal the presence of diffuse, spike-like or sheet-like structures 
called plumes \citep{1975spre.conf..651B, 1977SoPh...53..397A}, 
which subtend an angle of roughly $2\degree$ relative to Sun center at low altitude and 
expands super-radially with the coronal hole \citep{1997SoPh..175..393D}. 
Regions between these structures are termed as inter-plumes. 
From VUV spectroscopy, plumes are known to be denser and cooler than the 
surrounding inter-plume regions \citep[e.g.,][]{2006A&A...455..697W}, while 
spectral lines are observed to be broader in inter-plumes 
\citep[i.e.,][]{2000SoPh..194...43B,2000ApJ...531L..79G,2003ApJ...588..566T}. 
However differences in mass, momentum and 
energy flux in plumes and in inter-plumes are still not known precisely.
 
There are several theoretical models which describe the role of MHD waves in 
the acceleration of the fast solar wind in coronal holes 
\citep[see review by][and references therein]{2005SSRv..120...67O,2009LRSP....6....3C} 
and inter-plumes are often believed to be the primary site for this acceleration. 
It is further conjectured that these waves originate from the on-disk bright network regions 
\citep{1997ApJ...484L..75W,2000A&A...359L...1P,2000ApJ...531L..79G,2001A&A...380L..39B}. 
A number of studies \citep{1997ApJ...491L.111O, 2000ApJ...529..592O, 2001A&A...380L..39B, 
2005A&A...442.1087P} have reported detection of oscillations in the off-limb regions of polar 
coronal holes. All of these studies point to the presence of compressional waves, thought to
be slow magneto-acoustic waves 
\citep{1998ApJ...501L.217D, 2006A&A...452.1059O, 2007A&A...463..713O, 2009A&A...499L..29B}. 
On the other hand, evidence for Alfv\'{e}n waves propagating into the corona had been reported by
\citet{1998A&A...339..208B,2009A&A...501L..15B,2008A&A...483..271D,2009ApJ...691..794L} by studying
 the line width variations with height in polar coronal holes. Recent reports of  
detections of low-frequency ($<$ 5 mHz), propagating transverse motions 
in the solar corona \citep{2007Sci...317.1192T} (from coronagraphic observation) and chromosphere 
\citep{2007Sci...318.1574D} and their relationship with chromospheric 
spicules observed at the solar limb \citep{2007PASJ...59S.655D} with the 
Solar Optical Telescope aboard Hinode 
\citep{2007SoPh..243....3K} have widened interest in the subject. Recently \citet{2009Sci...323.1582J} have
 reported detection of torsional Alfv\'{e}nic motions associated with a large on-disk bright-point group. 
These waves are believed to be a promising candidate for the heating of the corona and acceleration of the solar wind
 \citep{1971ApJ...168..509B,2005ApJ...632L..49S}.

Furthermore, it has been suggested that the fast solar wind streams originate 
from coronal hole funnels and are launched by reconnection at network 
boundaries, \citep{2005Sci...308..519T}. 
Measurements of the outflow speed in the extended corona have been obtained with the Ultraviolet 
Coronagraph Spectrometer (UVCS) aboard SoHO 
\citep[e.g.,][]{2000SoPh..197..115A,2003ApJ...588..566T,2004A&A...416..749A,
2007A&A...472..299T}.
Some of these studies concluded that plumes have lower outflow speeds than inter-plume regions 
\citep{1997AdSpR..20.2219N,2000ApJ...531L..79G,2000A&A...353..749W,2000A&A...359L...1P,2003ApJ...588..566T,2007ApJ...658..643R} 
and, hence, may not contribute significantly to the fast solar wind, whereas some other theoretical 
and observational studies find higher outflow speeds in plumes than in inter-plume
regions for at least some altitudes above the photosphere 
\citep{1999JGR...104.9947C,2003ApJ...589..623G,2005ApJ...635L.185G}. 
These contradictory reports led to the debate on whether plumes or inter-plumes are the 
preferred source regions for the acceleration of the fast solar wind. 
This topic is highly debated and still open for further confirmation.

Recently, \citet{2009A&A...499L..29B} reported the detection of propagating slow 
magnetoacoustic waves with periods between 10~min and 30~min and speed $\approx75$~km~s$^{-1}$
 to $125$~km~s$^{-1}$\ above the limb of a polar coronal hole. 
In their study, the propagating disturbances which are due to radiance perturbations are seen from the limb region up to $\approx100\arcsec$ 
above the limb. There is no discernible acceleration or deceleration of any individual feature as 
it propagates. In that study, the oscillations were detected in the two spectral lines of 
Ne~{\sc viii}~770~\AA\ and Fe~{\sc xii}~195~\AA\ observed with the Solar  
Ultraviolet Measurements Of Emitted Radiation 
\citep[SUMER,][]{1995SoPh..162..189W} aboard the Solar and Heliospheric  
Observatory (SoHO) and with the EUV imaging spectrometer  
\citep[EIS,][]{2007SoPh..243...19C} aboard Hinode, respectively.

In this paper, we combine again the capabilities of SUMER and EIS to observe the on-disk, 
limb and far off-limb region of the coronal hole, to search for the origin of waves close to 
the Sun and study their propagating nature. The plan of the paper is as follows: in 
section~\ref{sec:obs}, the observations
acquired for this study and the data reduction techniques are outlined. In section~\ref{sec:result}
 results of the present study are presented with the distance-time radiance map analysis, power
 series analysis and time-delay analysis. A discussion of the observational
results and a comparison with similar results
are taken up in  section~\ref{sec:discussion} and finally conclusions are drawn in
 section~\ref{sec:conclusion}.

\section{Observations}
\label{sec:obs}
\subsection{Data}
\begin{figure}[htbp]
\centering
\includegraphics[angle=0, width=16cm]{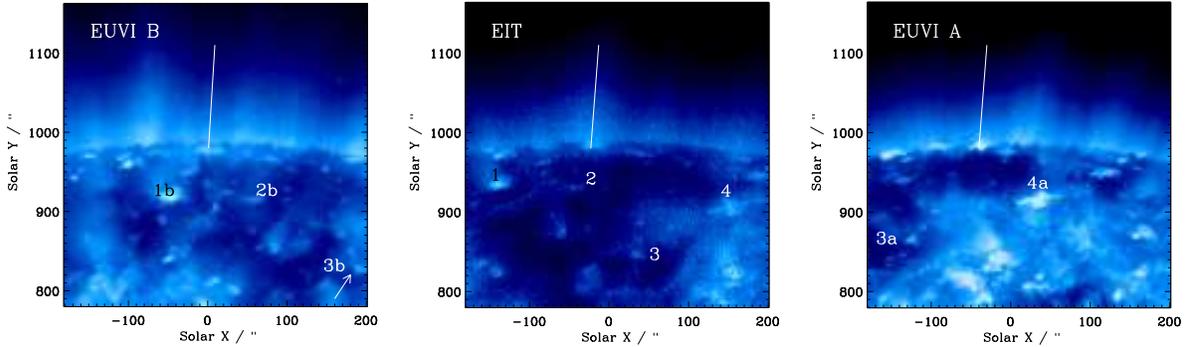}
 \caption{EUVI/STEREO and EIT/SoHO images in the 171 \AA bandpass taken around 
19:00 UT on $13^{th}$ November 2007. The solid line shows the plume axis as obtained by fitting the
 positions of the radiance maxima of horizontal cuts on the EIT image. Separation
 between the two STEREO spacecraft was $40\degree$ ($20\degree$ with Earth/SoHO). The different
 angular scale (km/\arcsec) was taken into consideration. Numbers from 1 to 4 on the EIT image identify
 reference elements that can be recognized on the EUVI-B image (suffix b) and/or on the EUVI-A
 image (suffix A).}
 \label{fig:stereo}
\end{figure}

The data analyzed here were obtained on 13$^{th}$ November  
2007 during a Hinode/SUMER joint observing campaign as part of the 
Hinode Observing Programme (HOP)~45/Joint Observing program (JOP)~196. The data 
consist of time series taken by SUMER and EIS in the north polar coronal hole.   
For SUMER, the $1\arcsec \times 120\arcsec$ slit was centered on the limb 
and spectral profiles of the Ne~{\sc viii} 770~\AA, O~{\sc iv} 790~\AA\ and 
S~{\sc v} 786~\AA\ were acquired from 19:13 to 22:15 UTC with an average cadence 
of 18.12~s in sit and stare mode. The
exposure time was 18~s and a total of $600$ time-frames were obtained during the 
observation. 
For EIS, the $40\arcsec$ wide slot was used to obtain  
$40\arcsec\times 512\arcsec$ images in several spectral lines in the wavelength ranges of
170 to 210 \AA\ and 250 to 290 \AA\ with spatial resolution of $1\arcsec$ pixel$^{-1}$ 
over the time interval from 18:20 to 23:50 UTC.
The exposure time was 45~s with an effective cadence of $\approx47$~s. A total of $420$ time-frames
were obtained during the observation. 
Before the start of the temporal series, raster images were obtained with SUMER and EIS in 
order to co-align and to provide context. 
During the observation, the EIS slot covered the quiet Sun south of the coronal hole as well as the on-disk and 
off-limb parts of the hole (see Figure~\ref{fig:map}). Table ~\ref{tab:lines} lists the 
emission lines included in this study from EIS and SUMER, their formation temperature, 
and the location where the radiance maximum is observed. Top left panel of 
Figure~\ref{fig:map} shows the location of the different slits on an EIT image taken in the 
Fe~{\sc xii}~195~\AA\ passband. 
The rectangular box marks the location of the EIS slot while the dashed line gives the
location of the SUMER slit. The radiance variation along the solar-X at 
solar-Y$\approx1000\arcsec$ is over-plotted as a white line in arbitrary units and allows us to 
identify the locations of plume and inter-plume regions within our field 
of view revealing that the SUMER slit is pointed within an inter-plume region 
while the EIS slot covers both plume and inter-plume regions. Bottom left shows the context
 raster obtained by EIS in Fe~{\sc xii} line whereas right panels show the context rasters
  in O~{\sc iv} (top) and Ne~{\sc viii} (bottom) spectral lines as obtained by SUMER. 
Figure~\ref{fig:stereo} corresponds to the images obtained by EUVI/STEREO
 \citep{2008SSRv..136...67H} for the same region. The angular separation between the two
 spacecraft, about 40~$\degree$, allows an estimate of the orientation of the plume.


\begin{table}[htb]
\caption{Emission lines observed with EIS and SUMER and position of the
respective limb brightening}
\begin{tabular}{cccc} 
\hline \hline
Ion & Wavelength (\AA) & log T$_{\rm max}$ (K) & Limb brightening \\
\hline
He~{\sc ii} & 256.32 & 4.9 & 985\arcsec \\
S~{\sc v} &   786.47 & 5.2 & 988\arcsec\\
O~{\sc iv} &  790.19 & 5.2 & 989\arcsec\\
Mg~{\sc vi} & 270.39 & 5.6 & 990\arcsec \\
Fe~{\sc viii} & 185.21 & 5.6 & 990\arcsec \\
Mg~{\sc vii} & 278.40 & 5.8 & 991\arcsec \\
Si~{\sc vii} & 275.35 & 5.8 & 993\arcsec \\
Ne~{\sc viii} & 770.42 & 5.8 & 993\arcsec\\
Fe~{\sc x} & 190.04 & 6.0 & 993\arcsec \\
Fe~{\sc xi} & 188.23 & 6.1 & --\\
Fe~{\sc xii} & 195.12 & 6.1 & 992\arcsec\\
\hline
\end{tabular}
\label{tab:lines}
\end{table}

\subsection{Data reduction and alignment} 
 
\begin{figure*}[htbp]
\centering
\includegraphics[width=16cm]{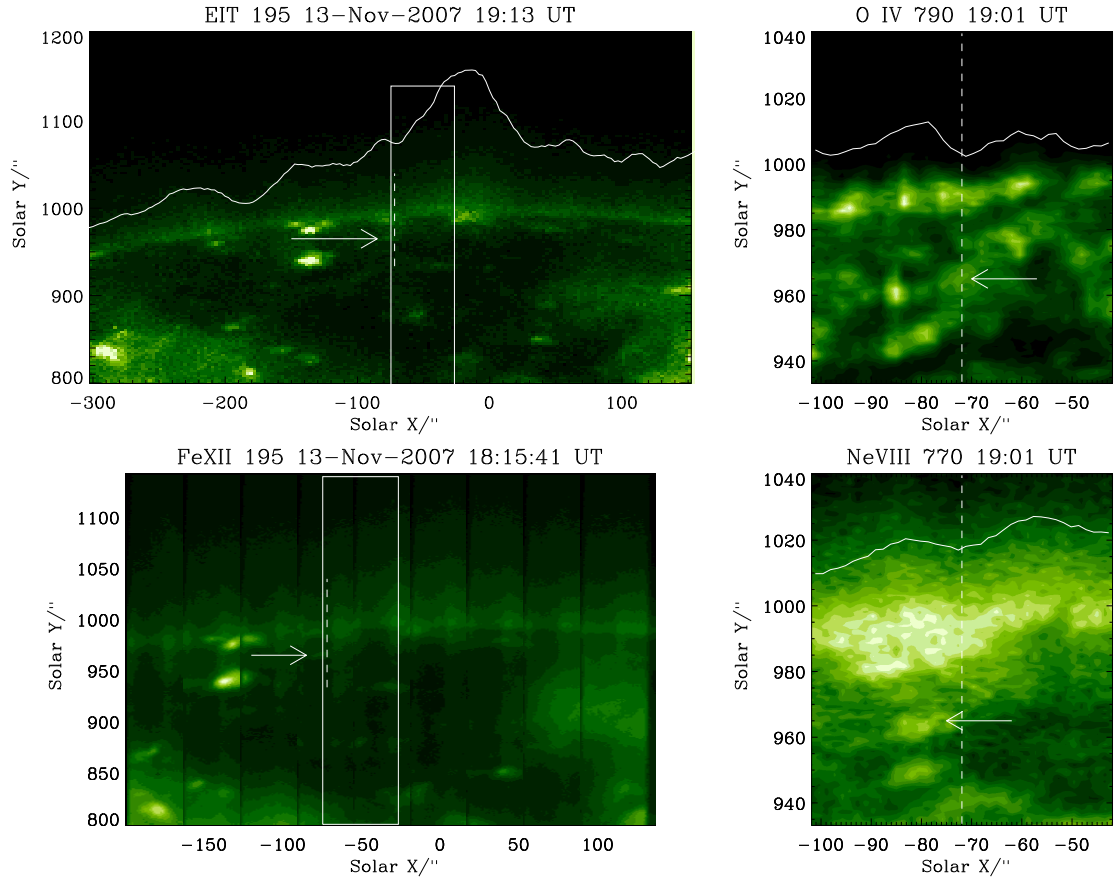}
\caption{Top left: The location of the different slits are over-plotted on the EIT image taken on $13^{th}$
 November 2007 in the passband dominated by the Fe~{\sc xii}~195~\AA\ line. The rectangular box
 marks the location of the EIS slot while the dashed line gives the
 location of the SUMER slit. The EIT radiance variation along the solar-X is
 over-plotted as a white line in arbitrary units at fixed solar-Y$\approx1000\arcsec$. This variation along
 the solar-X allows us to identify the location of plume and inter-plume regions. Bottom left: The EIS 
context raster taken in the same Fe~{\sc xii} line shows the location of both the EIS slot 
(rectangular box) and the SUMER slit (dashed line) during the sit-and-stare sequence. 
 Right: The SUMER context rasters taken in the O~{\sc iv} (top) and Ne~{\sc viii} (bottom) spectral
 lines. The continuous line gives the radiance variation along the 
solar-X in an arbitrary unit. In all panels, the arrow indicates the location of the 
bright region from where waves are presumably originating.}
\label{fig:map}
\end{figure*}

\begin{figure}[htbp]
\centering
\includegraphics[angle=90, width=14cm]{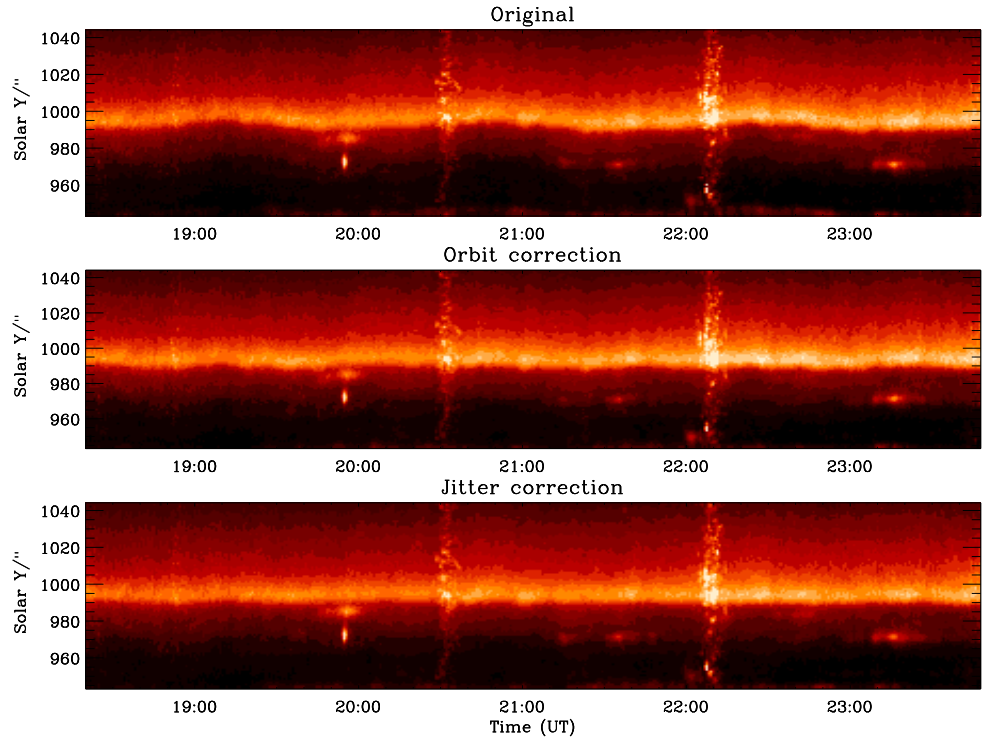}
\caption{Variation of radiance over time along the slice of corona recorded by both 
 SUMER and EIS. The 
 top panel shows the distance-time map of the Fe~{\sc xii} line as recorded by EIS. 
 Here the limb clearly shows the effect due
 to orbital variation and instrumental jitter. The middle panel shows the distance-time map 
 after orbit correction and the bottom panel shows the map after orbit and jitter correction 
 applied, showing the limb position to be much more stable in solar-Y 
as compared to the uncorrected data. In the panels, two visible vertical stripes around 20:35 
and 22:15 UTC are damaged data, probably due to SAA transits (another, less visible stripe is
present around 18:55 UTC).}
\label{fig:limb_jitter}
\end{figure}

\begin{figure}[htb]
\centering
\includegraphics[angle=90, width=10cm]{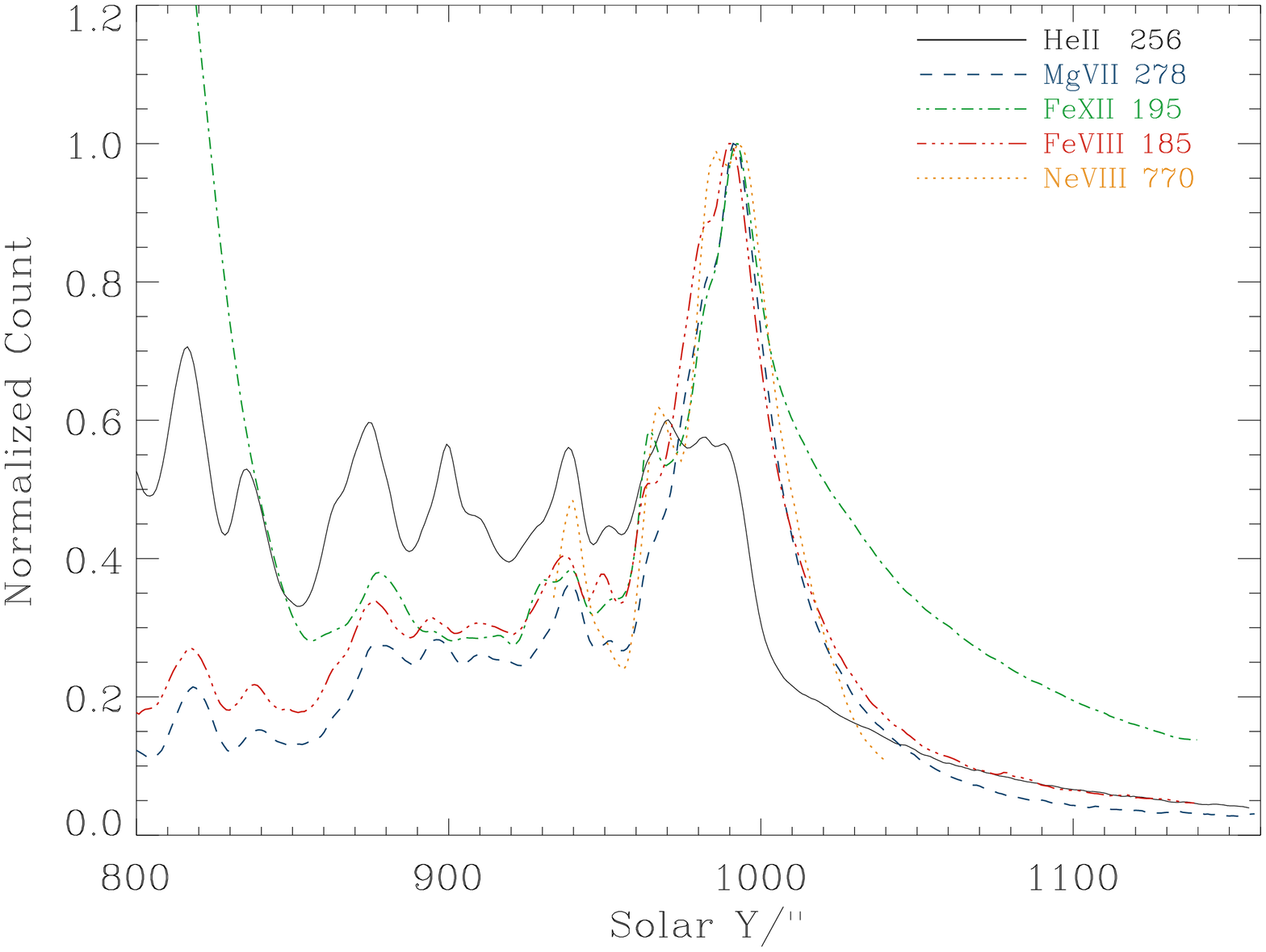}
 \caption{Time averaged radiance variations along the SUMER slit and EIS slot (solar-Y)
 at solar-X$\approx -72\arcsec$\ for different lines as labeled.}
 \label{fig:limb_var}
\end{figure}

All data have been reduced and calibrated with the standard procedures given in the SolarSoft  
(SSW)\footnote{\url{http://sohowww.nascom.nasa.gov/solarsoft/}} library.  
SUMER data were first decompressed, corrected for response inhomogeneities  
(flatfield), dead-time, local-gain and for geometrical distortion (de-stretch), using the most
 recent standard routines \citep[see][]{1997SoPh..170...75W,1999A&A...349..636T}. After these 
steps, data still showed a residual pattern from the micro-channel plate structure that was removed
 using a correction matrix obtained by first averaging all spectral images and then applying a
 low pass filter to the average. 
Single Gaussian fitting was used to retrieve the line amplitude, position and line widths of the
 SUMER spectral lines. Before fitting, a running average over three pixels along the slit and over
 three consecutive spectra was applied to improve the signal to noise ratio of the SUMER data. 
Line positions from the fitting are then converted into Doppler shifts by taking as reference
 the average over the disk part of the image. The observed line widths are corrected for the
 instrumental profile by applying a de-convolution function taking into account the order of
 diffraction and the slit width used during the observations (SolarSoft routine 
$con_{-}width_{-}funct_{-}4$).
The raw EIS data were processed by the standard SolarSoft programme 
$eis_{-}prep$ which helps in removing detector bias and dark current, hot pixels and 
cosmic rays and returns absolutely calibrated data.
The movement of the slot image on the detector due to thermal variations
along the orbit was corrected. The displacement in the dispersion (solar X) 
direction was obtained by measuring the position of the edge of the Fe~{\sc xii} 195 \AA\ 
slot image over time. The displacement in the Y direction is taken equal to 2.5
times that in the X direction \citep{2010Imada}. 
The validity of the latter assumption was verified by checking the limb Y
position vs. time. Finally, EIS data were corrected for the spacecraft jitter by using housekeeping data.
Figure~\ref{fig:limb_jitter} shows the variation of the radiance along a
vertical strip (about 100 pixels long, centered over the limb) at the overlap
position of the EIS slot with the SUMER slit as a function of time (x-t slice). The x-t 
slice are for data without any correction (top panel), with the orbital effects corrected
(middle), and with correction for both orbital and jitter effects (bottom). Data are also affected
by passage of the spacecraft through South Atlantic Anomaly (SAA). The regions of affected data
appear around 20:35 and 22:15 UTC and, less clearly visible, around 18:55 UTC. These affected portions are
 replaced by linear interpolation.
It should be finally noticed that for wavelet analysis, EIS data have been
binned over 5 and 9 pixels in the X and Y-directions, respectively.
This will further smooths out any residual jitter or orbital variation.
The data from the short wavelength detector were shifted in the Y-direction to compensate 
the wavelength dependent offset between the short and long wavelength detectors 
(Kamio \&\ Hara, private communication).
To align the different instruments, the SUMER context raster has been chosen as the reference.
Hence, the EIS context raster has been cross-correlated with the SUMER raster and  an
offset of $9\arcsec$\ in the E-W direction (solar X), and $-24\arcsec$\ in the N-S 
direction (solar Y) was found and corrected for. After the alignment, the pointing of the different
instruments are plotted in Figure~\ref{fig:map}. We have plotted the variation of 
radiance along the solar-Y at the overlap region of the SUMER slit and EIS slot at
solar-X $\approx-72\arcsec$, see Figure~\ref{fig:limb_var}.
This allows us to identify the location of the limb brightening in various
spectral lines and their radiance fall-off in the off-limb region. 
We set the limb position at solar-Y$\approx985\arcsec$, as 
identified from the limb brightening of He~{\sc ii} in Figure~\ref{fig:limb_var}.
The radiance peaks of the He~{\sc ii} line for the disk part of the coronal 
hole are used to identify bright locations (presumably the footpoint of the coronal funnels).

\section{Results}
\label{sec:result}
In this section we will present detailed analysis of the sit-and-stare 
observation as recorded by EIS and SUMER. From the context image, 
Figure~\ref{fig:map}, it is clear that the SUMER slit is pointing to an 
inter-plume region and the EIS slot covers both plume and 
inter-plume region, overlapping with the SUMER slit. 
Thus, in the inter-plume region, EIS and SUMER will both provide information 
whereas in the plume region only EIS observations are available. 
Analysis of the radiance 
maps (x-t slices), wavelet analysis and correlation analysis will be  performed 
at the two locations as described in the following three subsections.

\subsection{Radiance x-t slices}
Maps of radiance along the slit vs. time (distance-time map or x-t slices) were 
 built using the SUMER Ne~{\sc viii} integrated line radiance and the EIS 
Fe~{\sc xii} radiance averaged over $5\arcsec$ in the X direction at the 
position overlapping  with the SUMER slit (which covers 
the inter-plume region). The resulting maps were then smoothed over $\approx3$~min and 
the background trend of $\approx20$~min has been subtracted from each solar-Y 
pixel along time. 
In general, similar procedures are applied while doing the Fourier or wavelet 
analysis of a time series. In the x-t slices, the presence of alternate bright 
and dark regions indicate the presence of oscillations. Moreover, diagonal or 
slanted radiance enhancements are signature of propagating 
disturbances. Thus, from such maps it becomes
possible to estimate periods and projected propagation speeds 
\citep[see e.g.][]{1998ApJ...501L.217D}.
 
\begin{figure*}[htb]
 \centering
\includegraphics[width=14cm]{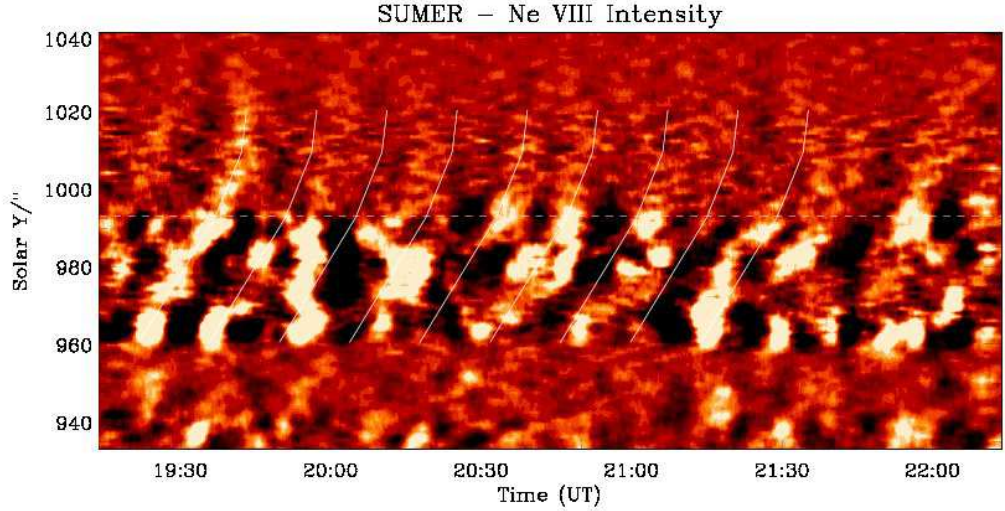}
\caption{Enhanced distance-time (x-t) map of radiance (along solar-Y) variation 
at solar X$\approx -72\arcsec$ as recorded by SUMER in the Ne~{\sc viii} 
spectral line on $13^{th}$ November 2007. Here the slit covers the on-disk, 
limb and off-limb region of the polar coronal hole and it is positioned in the 
inter-plume region. 
The slanted lines correspond to the disturbances propagating outward with 
increasing speed. 
The dashed horizontal line indicates the position of the limb 
brightening in Ne~{\sc viii}.
In the on-disk region the disturbance propagates with a speed of ($25~\pm~1.3$)~km~s$^{-1}$, 
increasing to ($38~\pm~4.5$)~km~s$^{-1}$ close to the limb and to about ($130~\pm~51$)~km~s$^{-1}$ in
 the off-limb region.
The periodicity is in the range of $\approx 14$~min to 20~min as also obtained from the 
wavelet analysis, see Figures~\ref{fig:ne967} \& ~\ref{fig:ne1020}. }
\label{fig:xt_ne8}
\end{figure*}

Both the SUMER slit and the EIS slot are centered on the solar limb and, hence, 
cover the region on-disk as well as off-limb. 
As the observed region is near to the pole, the effect of solar rotation is 
very small and amount to less than about $3\arcsec$ per hour at 
$100\arcsec$ below the limb.
We first concentrate our attention to the inter-plume location, around 
solar-X$\approx-72\arcsec$, probed by both instruments.
Figure~\ref{fig:xt_ne8} shows the x-t map of the radiance of the Ne~{\sc viii} 
spectral line, where the presence of slanted bright and dark region is clearly 
visible. 
A disturbance appears from the on-disk bright region around solar-Y $\approx967\arcsec$\ 
(see Figure~\ref{fig:map} and \ref{fig:limb_var}), and propagates towards 
the limb. The signature of oscillations are very strong in this bright region. 
No signature of propagation is visible below solar-Y$\approx967\arcsec$\ 
(the SUMER slit covers down to solar-Y$\approx930\arcsec$). 
Hence the assumption that this bright region is the source of these propagating 
disturbances is justified (Figure~\ref{fig:xt_ne8}). 
The speed of propagation measured from the slope of the
enhanced slanted radiance stripes is ($25~\pm~1.3$)~km~s$^{-1}$. 
This average speed is measured up to solar-Y$\approx992\arcsec$\ which is very close to the 
region of limb brightening (see also Figure~\ref{fig:limb_var}). 
As the propagation reaches the limb, its speed increases and the enhanced features become 
more vertical. 
Up to solar-Y$\approx1010\arcsec$, the measured speed is ($38~\pm~4.5$)~km~s$^{-1}$. This
change in speed is a clear signature of acceleration of the propagating disturbance. 
Furthermore, when this propagation reaches beyond the limb brightening height \ie in the corona, 
its speed further increases to ($130~\pm~51$)~km~s$^{-1}$ and the
propagation is seen up to solar-Y$\approx1020\arcsec$. Beyond this height the signature of 
propagation becomes very poor, most likely due to the low signal. 
The periodicity of the fluctuations is $\approx14$~min to 20~min. 
In Figure~\ref{fig:xt_ne8}, the over-plotted white lines follow the slope of the enhancements 
and are plotted with 
a periodicity of $\approx14$~min. It can be seen that in some
places the over-plotted white line do not coincide with the enhanced lanes but it is
nevertheless parallel to it. 
This suggests that even if the periodicity changes within a certain range, the propagation speeds 
are fairly uniform.
There is no clear evidence of propagating disturbances 
(in terms of velocity fluctuations) in the line-of-sight (LOS) velocity (obtained from Doppler 
shift) x-t map, although periodicities are revealed by the wavelet analysis (see next section).

\begin{figure}[htbp]
 \centering
\includegraphics[width=7.5cm]{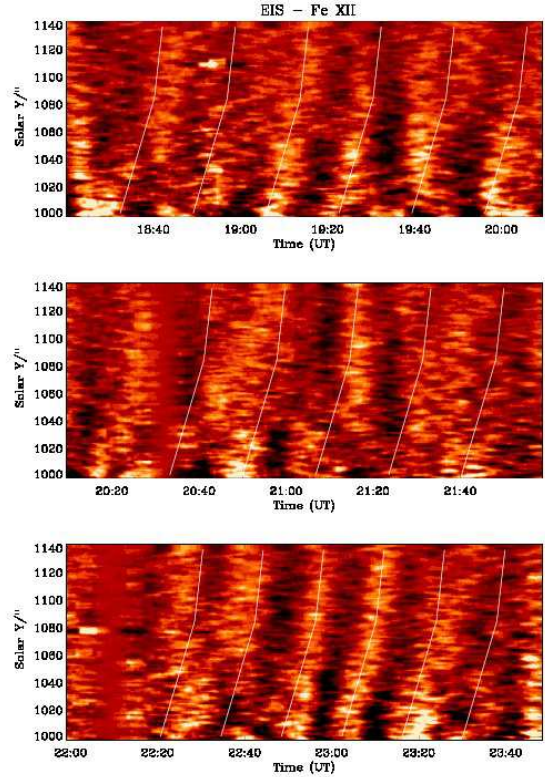}
\caption{Enhanced x-t map of radiance variation along solar-Y at
 solar-X$\approx-72\arcsec$ as recorded by EIS in Fe~{\sc xii} on 
$13^{th}$ November 2007. The height range shown here covers the near off-limb and far off-limb region 
of the polar coronal hole and corresponds to the inter-plume region. 
The slanted lines correspond to the disturbances propagating outward with increasing speed. 
In the near off-limb region the disturbance propagates with speed of ($130~\pm~14$)~km~s$^{-1}$, 
and accelerates to ($330~\pm~140$)~km~s$^{-1}$ in the far off-limb region. The periodicity
is in the range of
$\approx 15$~min to 18~min as obtained from the wavelet analysis, see Figure~\ref{fig:fe1020}.}
\label{fig:xt_fe12}
\end{figure}

Figure~\ref{fig:xt_fe12} shows the radiance x-t map of the 
Fe~{\sc xii}~195 \AA\ line at the same 
location (solar-X $\approx-72\arcsec$, inter-plume region). 
The image was processed as described above.  
The analysis of the x-t map over the coronal hole, reveals that there is no clear signature of
propagating disturbances in the on-disk coronal hole region. These alternate bright and dark 
regions are clearly visible only around the limb and far off-limb, hence only these regions are 
plotted in Figure~\ref{fig:xt_fe12}. 
In this map, the propagating disturbances are visible from solar-Y$\approx1000\arcsec$\ up to the 
upper end of the EIS slot at solar-Y$\approx1140\arcsec$. 
In this inter-plume region, the disturbance propagates with an average speed of
($130~\pm~14$)~km~s$^{-1}$ from solar-Y$\approx1000\arcsec$\ to solar-Y$\approx1085\arcsec$\
increasing to an average speed of ($330~\pm~140$)~km~s$^{-1}$  up to solar-Y$\approx1135\arcsec$, 
clearly showing signatures of acceleration. The periodicity of the
fluctuations is in the range of $\approx$ 15~min to 18~min. 
Here the over-plotted white lines gives the slope of the enhancements
and are plotted with a periodicity of $\approx17$~min. It can be again seen here that in some
places the over-plotted white line do not coincide with the enhanced lanes but is nevertheless 
parallel to it as was discussed for the results from the Ne~{\sc viii} line.\\
  

Summarizing, observations in the Ne~{\sc viii} spectral line reveal a propagating disturbance 
originating in a bright region in the on-disk coronal hole that starts propagating towards the 
limb region with speed ($25~\pm~1.3$)~km~s$^{-1}$. Near the limb the speed 
increases to ($38~\pm~4.5$)~km~s$^{-1}$ and reaches ($130~\pm~51$)~km~s$^{-1}$ in the off-limb region. 
A similar speed is measured at the same height in the Fe~{\sc xii} spectral line by EIS. 
Hence, both instruments see approximatively the same speed in the same region, which is different 
from the result reported by \cite{2009A&A...499L..29B} for a likely plume region, where they find 
different propagation speeds in different lines. 
Further off-limb, the speed of the propagating disturbance reaches ($330~\pm~140$)~km~s$^{-1}$
as seen by EIS in the Fe~{\sc xii} spectral line. 
Overall, acceleration of propagating
disturbances are observed from on-disk to far off-limb in an inter-plume region simultaneously
by two different instruments on different satellites.\\

\begin{figure}[htbp]
 \centering
\includegraphics[width=7.5cm]{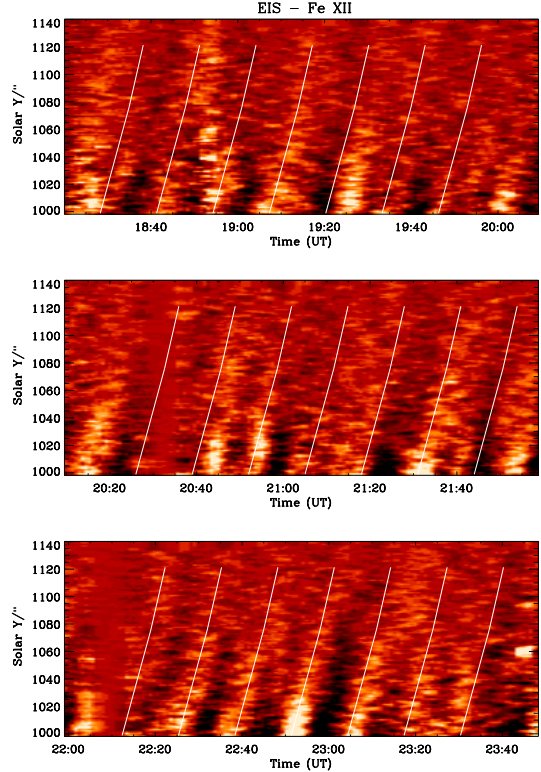}
\caption{Enhanced x-t map of radiance variation along solar-Y at
 solar-X$\approx-39\arcsec$ as recorded by EIS in Fe~{\sc xii} on 
$13^{th}$ November 2007. The height range shown here covers the near off-limb and far off-limb region of 
the polar coronal hole and falls in the plume region. The slanted lines corresponds to the disturbances 
propagating outward with nearly constant speed. 
In the near off-limb region the disturbance propagates with speed of ($135~\pm~18$)~km~s$^{-1}$, and accelerates 
 to ($165~\pm~43$)~km~s$^{-1}$ in the far off-limb region. The periodicity
 is in the range of
 $\approx 15$~min to 20~min as obtained from the wavelet analysis, see Figure~\ref{fig:fe1030}. }
\label{fig:xt_fe12p}
\end{figure}

The EIS slot also covers part of a plume and the position around
 solar-X$\approx-39\arcsec$ is selected to represent this region. 
 The processed x-t radiance map is plotted in 
Figure~\ref{fig:xt_fe12p}. Also in this case,  
there is no clear signature of propagating disturbances in the on-disk coronal hole region. 
These alternate bright and dark regions are clearly visible only at the limb and off-limb, hence 
again only these regions are plotted in Figure~\ref{fig:xt_fe12p}. 
In this map, the propagating disturbances are visible from solar-Y$\approx1000\arcsec$ up to 
solar-Y$\approx1120\arcsec$. 
In this plume region the disturbance propagates with a speed of
($135~\pm~18$)~km~s$^{-1}$ from solar-Y$\approx1000\arcsec$ to solar-Y$\approx1075\arcsec$ and 
with ($165~\pm~43$)~km~s$^{-1}$ up to solar-Y$\approx1120\arcsec$. 
Beyond this height the map becomes diffuse and there is no clear 
signature of propagation. Although this may be interpretated in terms of wave
dissipation, this lack of signature at greater heights may be simply due to
merging with the background signal.

\subsection{Analysis of oscillations}
\begin{figure*}[htbp]
\centering
\includegraphics[angle=90, width=8cm]{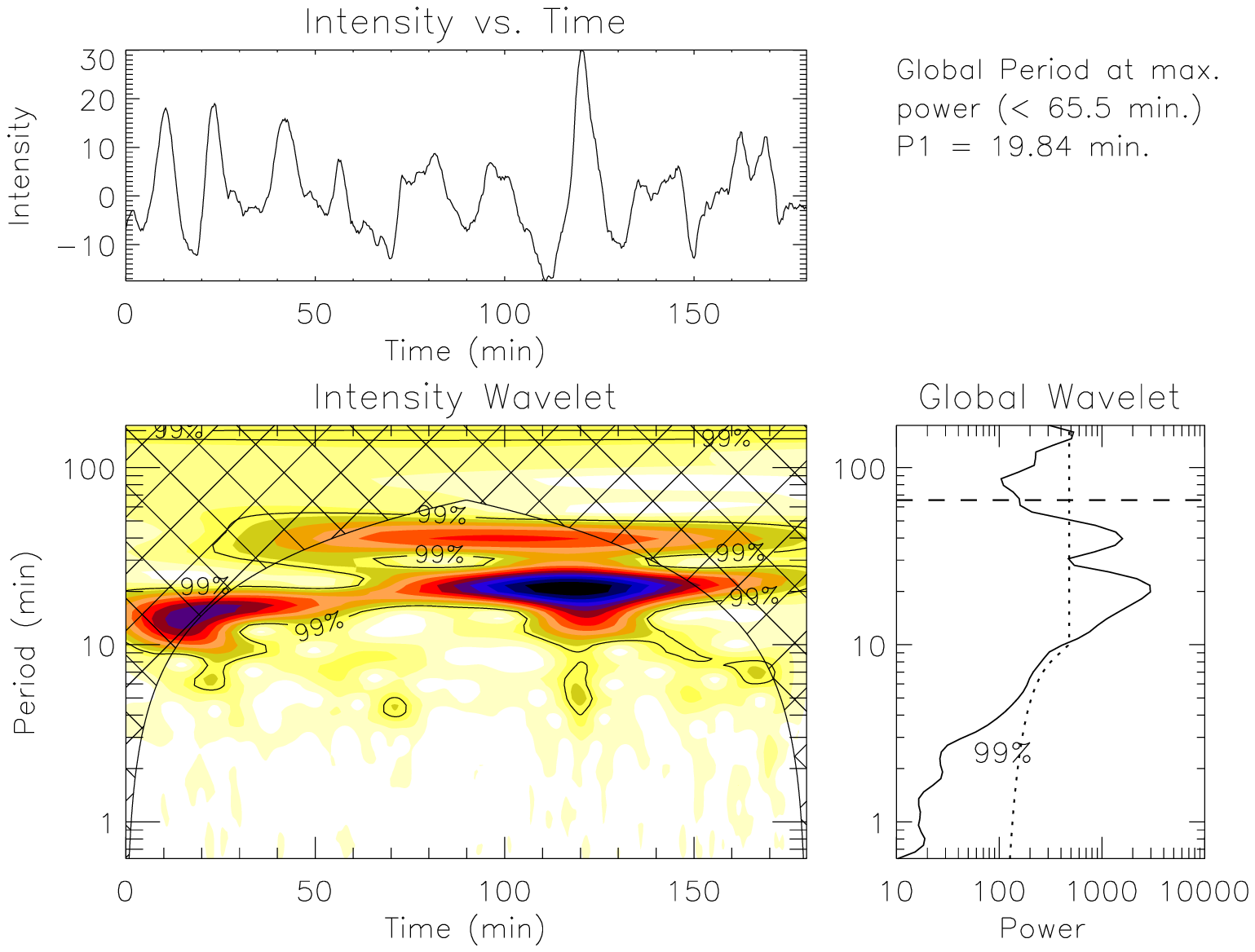}
{\includegraphics[angle=90,width=8cm]{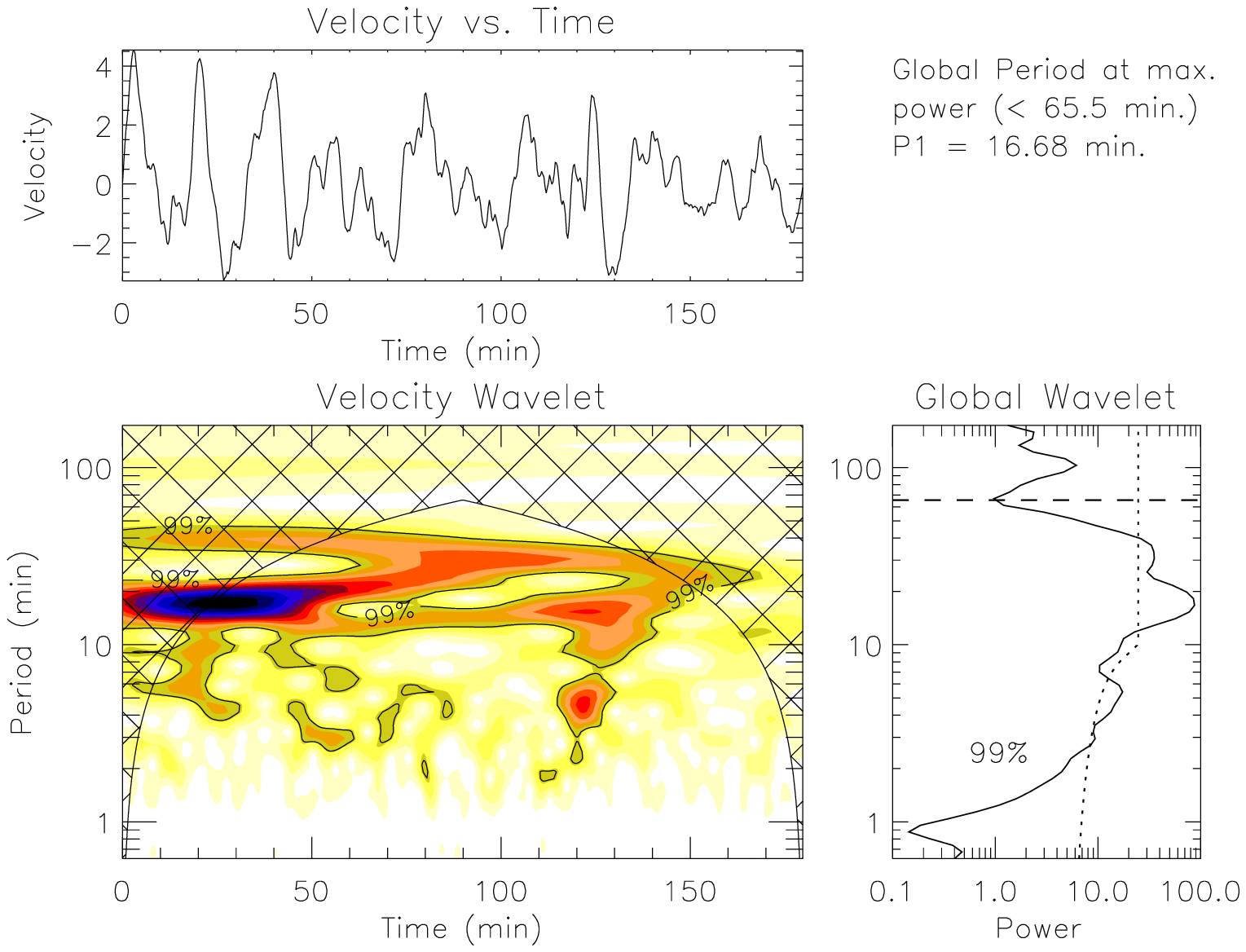}}
 \caption{The wavelet result for the on-disk location at
 solar-Y$\approx 967\arcsec$ and solar-X$\approx -72\arcsec$ in Ne~{\sc viii} radiance
 (left side) and velocity (right side). In each set, the
 top panels show the relative (background trend removed) radiance/velocity 
 smoothed over 3~min. Bottom left panels show the color inverted wavelet power
 spectrum with $99~\%$ confidence level contours while
 bottom right panels show the global (averaged over time) wavelet power spectrum with $99~\%$
 global confidence level drawn. 
 The period P1 at the location of the maximum in the global wavelet spectrum is 
 printed above the global wavelet spectrum.}
 \label{fig:ne967}
\end{figure*}

\begin{figure*}[htbp]
\centering
\includegraphics[angle=90, width=8cm]{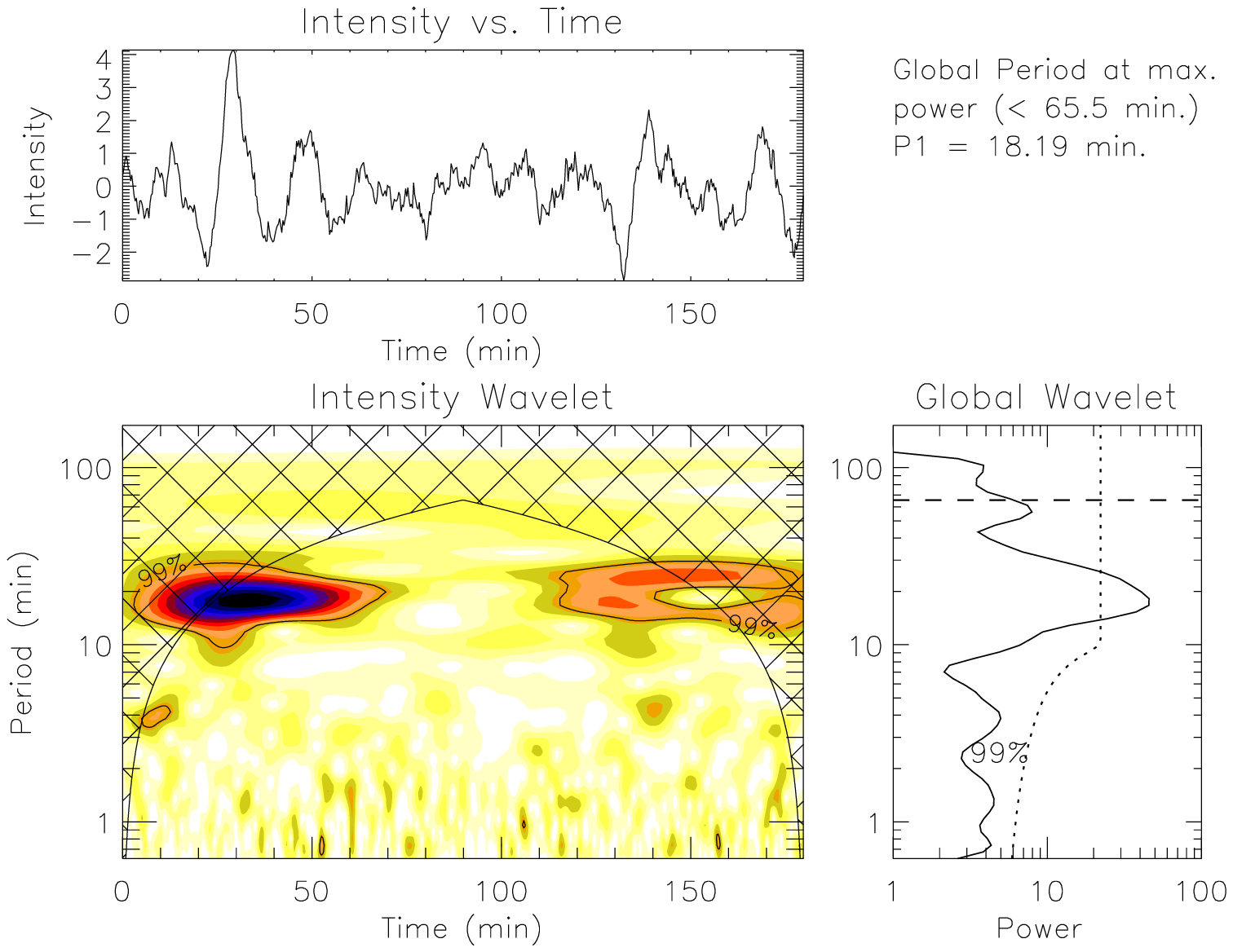}
{\includegraphics[angle=90, width=8cm]{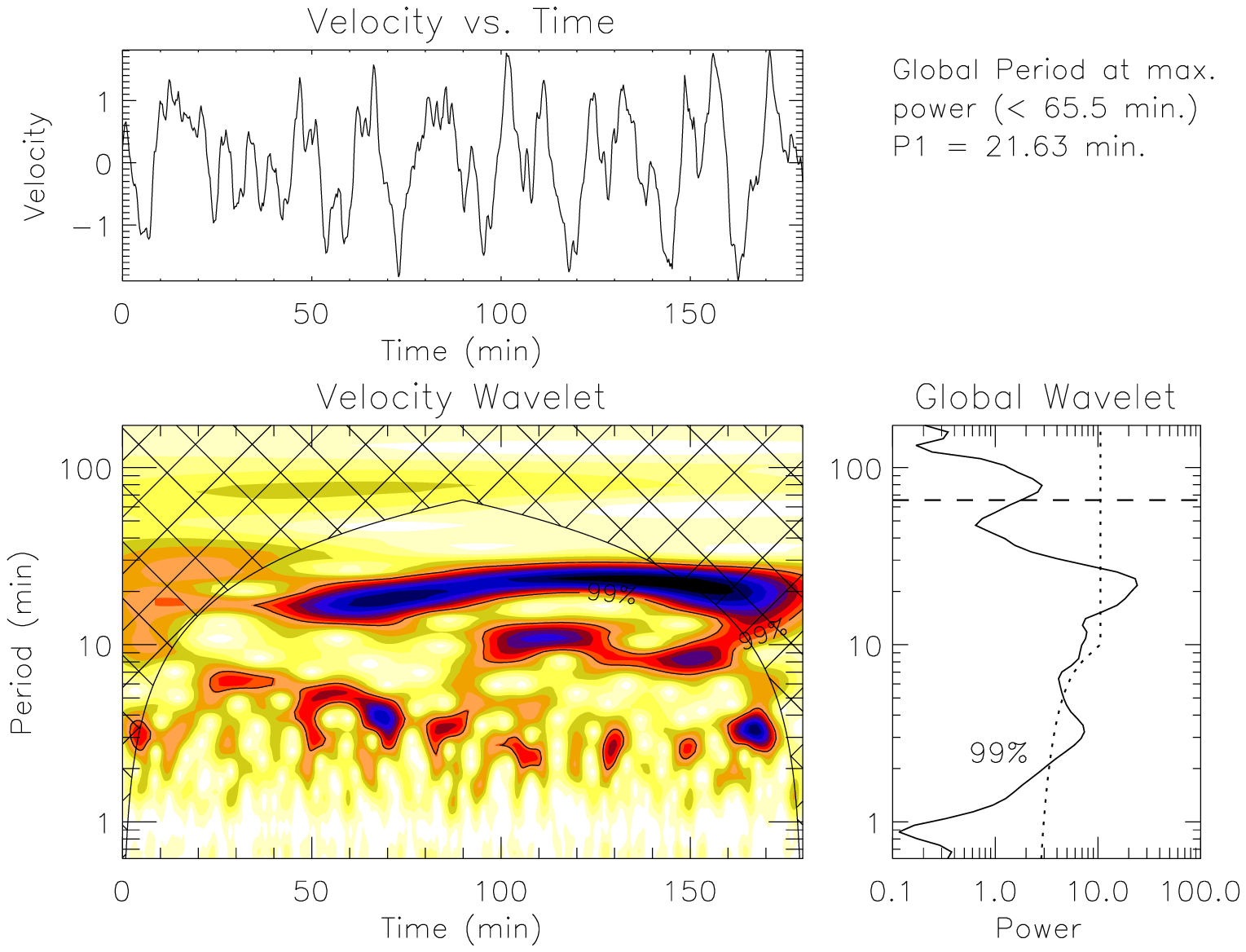}}
 \caption{The wavelet analysis results corresponding to solar-Y$\approx1020\arcsec$ in the Ne~{\sc viii}
 radiance (left side) and in velocity (right side) at solar-X$\approx-72\arcsec$
(inter-plume region). See the caption of Figure~\ref{fig:ne967} for a description of the different panels.}
 \label{fig:ne1020}
\end{figure*}

\begin{figure*}[htbp]
\centering
\includegraphics[angle=90, width=8cm]{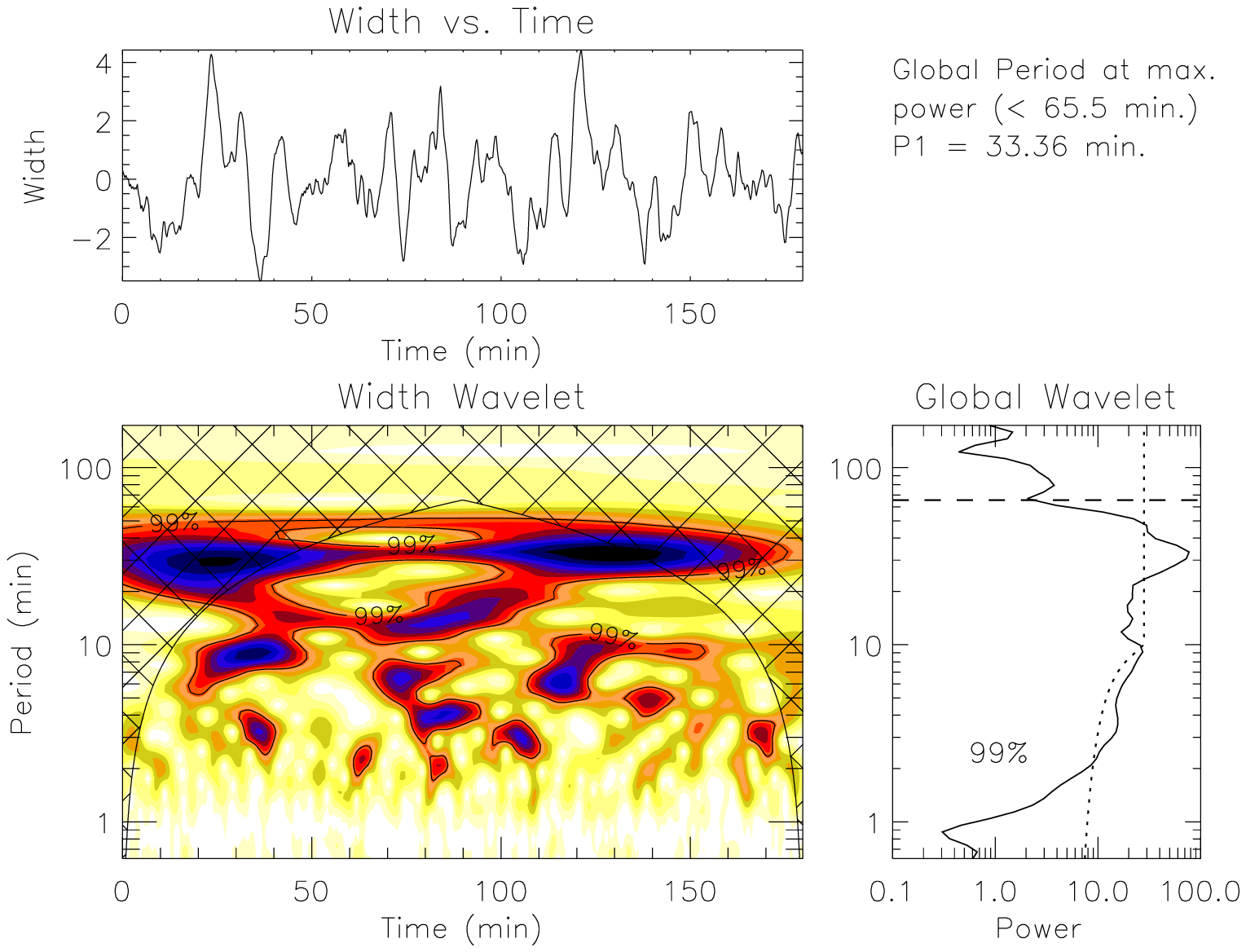}
{\includegraphics[angle=90, width=8cm]{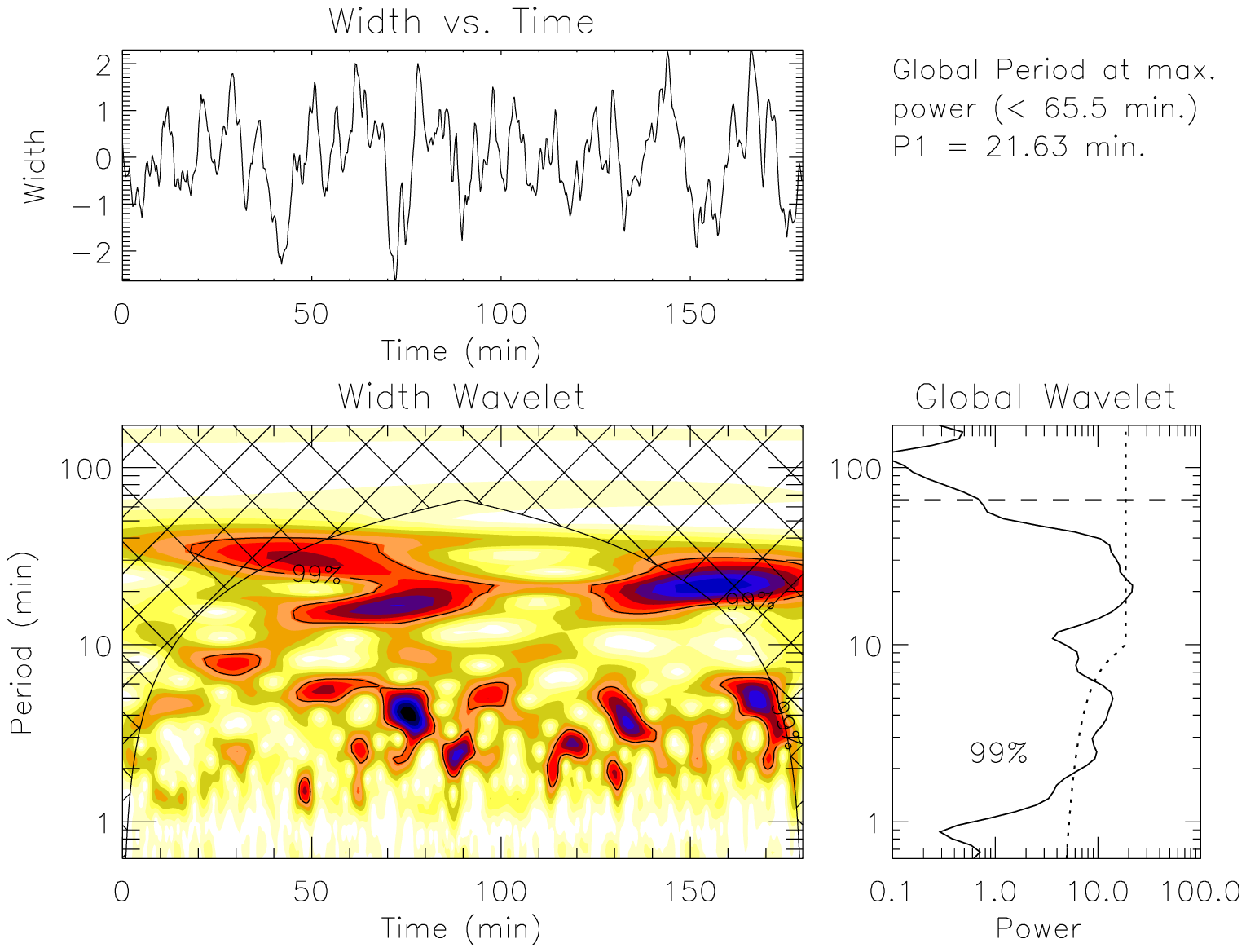}}
 \caption{The wavelet analysis results for the oscillations in Doppler width of the Ne~{\sc viii}
 line at solar-Y$\approx967\arcsec$ (left side) and at solar-Y$\approx1020\arcsec$ (right side) obtained
 at solar-X$\approx-72\arcsec$ (inter-plume region). See the caption of Figure~\ref{fig:ne967}
for a description of the different panels.}
 \label{fig:wdt_ne}
\end{figure*}

\begin{figure*}[htbp]
\centering
\includegraphics[angle=90, width=8cm]{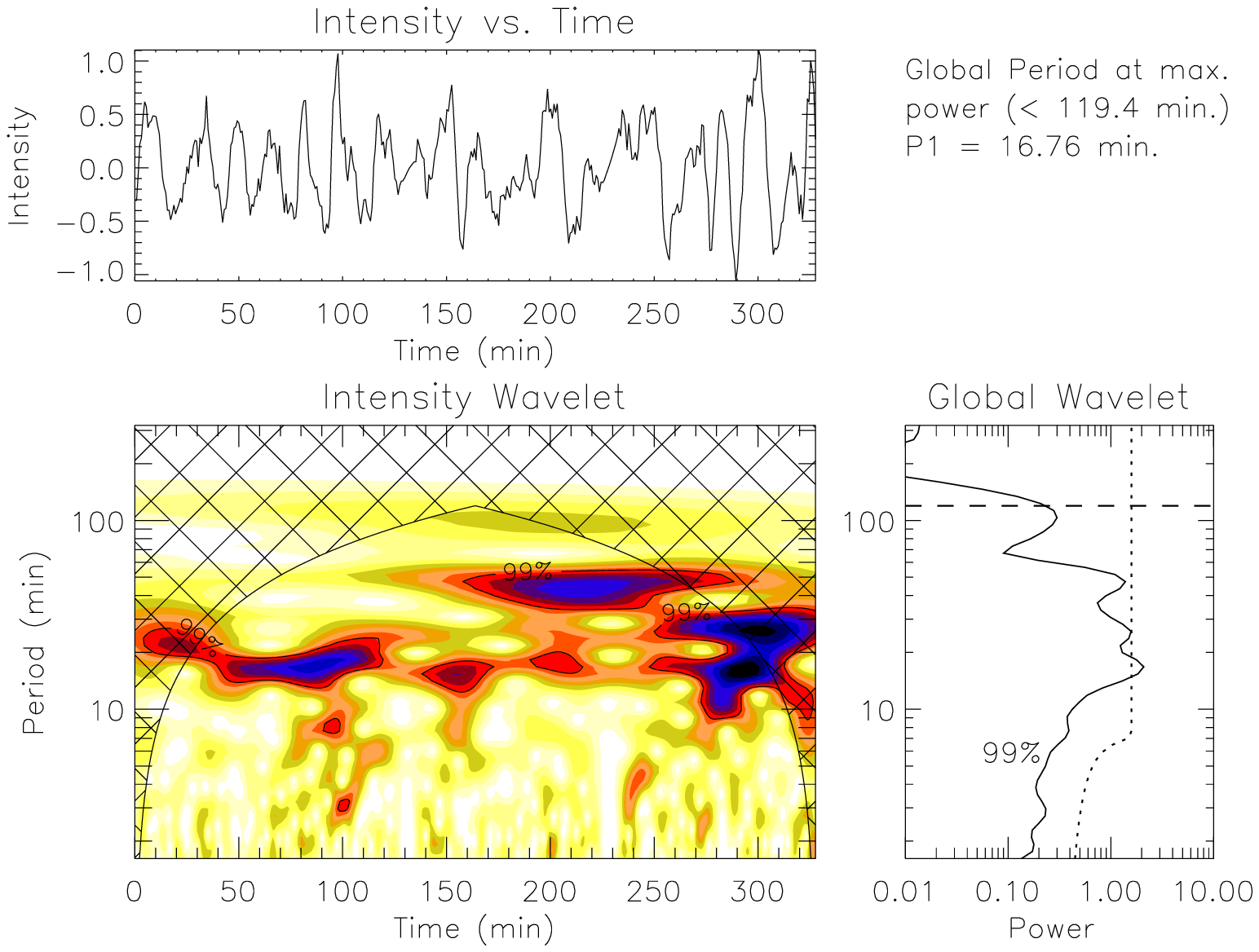}
{\includegraphics[angle=90, width=8cm]{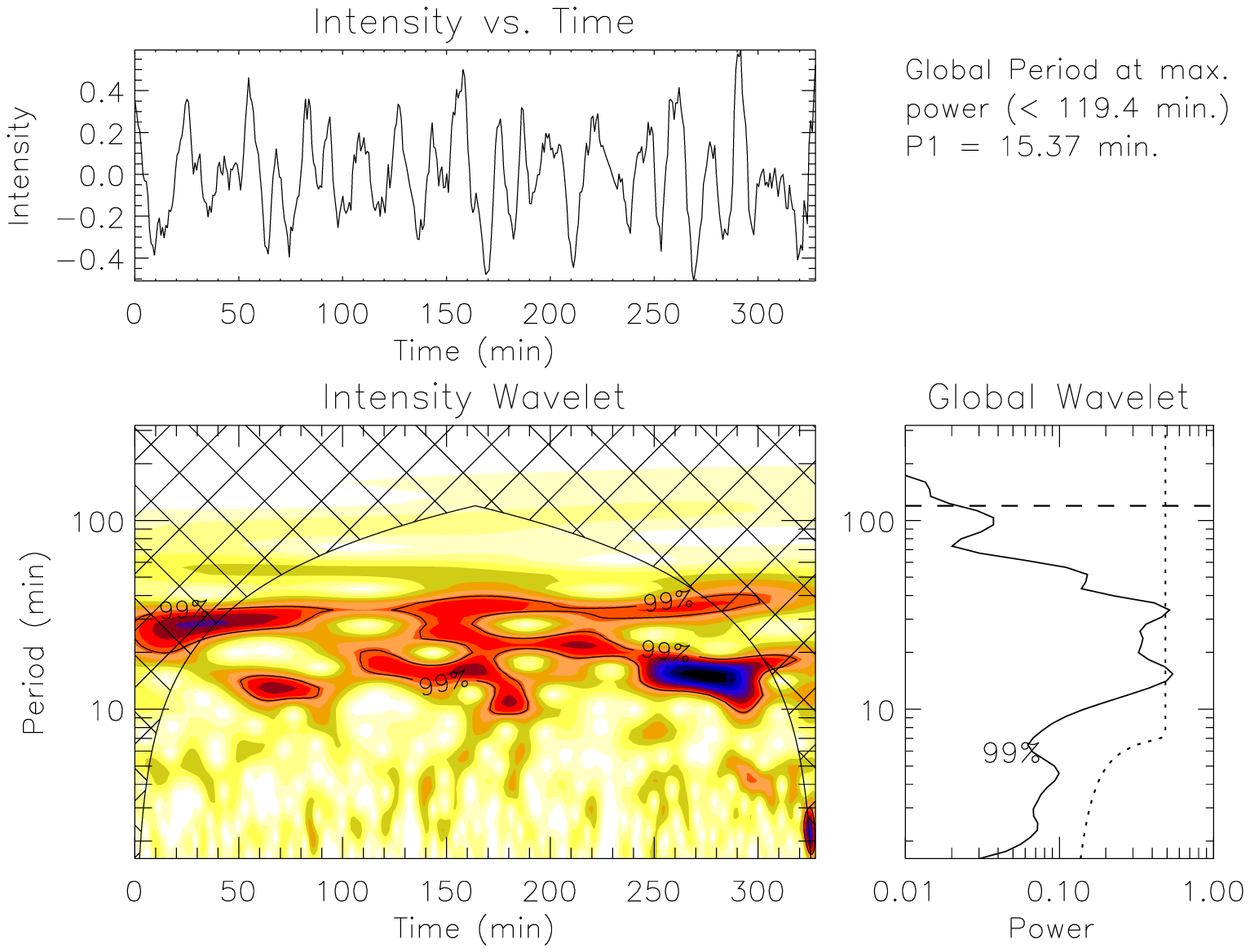}}
 \caption{The wavelet analysis results corresponding to the Fe~{\sc xii}
 radiance at solar-Y$\approx1020\arcsec$ (left side) and at solar-Y$\approx1120\arcsec$ (right side) at
 solar-X$\approx-72\arcsec$ (inter-plume region). See the caption of Figure~\ref{fig:ne967} for a
 description of the different panels.}
 \label{fig:fe1020}
\end{figure*}

\begin{figure}[htbp]
\centering
\includegraphics[angle=90, width=8cm]{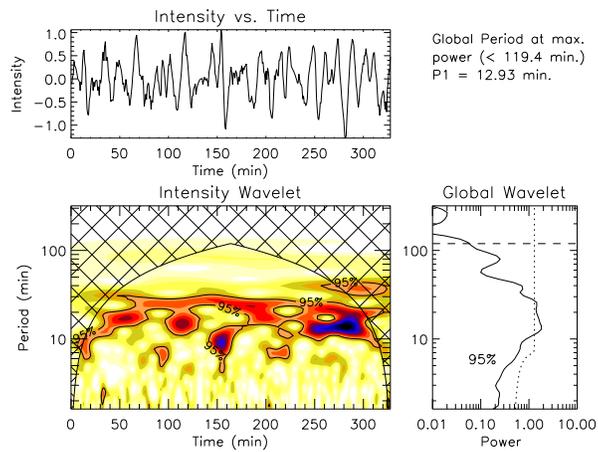}
 \caption{The wavelet analysis results corresponding to the Fe~{\sc xii}
 radiance at solar-Y$\approx1030\arcsec$ and solar-X$\approx-39\arcsec$ (plume region).
 See the caption of Figure~\ref{fig:ne967} for a description of the different panels. In this case,
confidence contours are drawn at $95~\%$ level.}
 \label{fig:fe1030}
\end{figure}

In this subsection, in order to study the detailed properties of the propagating disturbances as seen in the 
enhanced radiance x-t maps (Figures~\ref{fig:xt_ne8},~\ref{fig:xt_fe12} and~\ref{fig:xt_fe12p}), we 
make use of wavelet analysis and focus on individual locations in the on-disk  and off-limb corona.
The full SUMER time series is used to detect oscillations 
 in the radiance, Doppler shift as well as in the Doppler width of 
the Ne~{\sc viii} spectral line at several locations. 
The bright location is identified on-disk using the maximum radiance seen in
He~{\sc ii} $256$~\AA\ by EIS. At several off-limb
locations there is sufficient signal to noise to detect oscillations with a
high confidence level. Figures~\ref{fig:ne967},~\ref{fig:ne1020},~\ref{fig:wdt_ne} \& 
~\ref{fig:fe1020} show examples of oscillations measured in the polar region at fixed 
solar-X$\approx-72\arcsec$ (which corresponds to the inter-plume region) and at several solar-Y 
locations: $\approx967\arcsec$ (on-disk), $\approx1020\arcsec$ (off-limb, but close to the limb) 
and $\approx1120\arcsec$ (far off-limb), as mentioned in the figure caption.  
On the other hand, Figure~\ref{fig:fe1030} shows oscillation measured at
solar-X$\approx-39\arcsec$ (which corresponds to the plume region) and 
solar-Y$\approx1030\arcsec$ (off-limb).
In these figures the top panel shows the variation of the radiance (hereafter the term radiance will be 
used for trend subtracted integrated line radiance) with time. 
Details on the wavelet analysis, which provides information on the temporal variation of a signal, are 
described in \citet{1998BAMS...79...61T}. 
For the convolution with the time series in the wavelet transform, the Morlet function is choosen. 
The oscillations shown in the upper panel had their
background trend removed by subtracting from the original time series a $100-$point
($\approx30$~min) and $35-$point ($\approx30$~min) running average for SUMER and EIS data, respectively. 
In the wavelet spectrum, the cross-hatched regions are locations where estimates of oscillation period 
become unreliable which is called as cone-of-influence (COI). 
As a result of the COI, the maximum measurable period is shown by a horizontal dashed line in the
 global wavelet plots, which are obtained by taking the mean over the wavelet time domain. This global
wavelet is very similar to the Fourier transform as both are giving the distribution of power
 with respect to period or frequency. Whenever the Fourier spectrum is smoothed, it approaches the global
 wavelet spectrum.
The period at the location of the maximum in the global wavelet spectrum is 
printed above the global wavelet spectrum. 


From the x-t map analysis (Figure~\ref{fig:xt_ne8}) 
it was seen that outward propagating disturbances at the inter-plume location 
(solar-X$\approx-72\arcsec$) originate from a bright on-disk region  
at solar-Y$\approx967\arcsec$.
As this region has been covered by
the SUMER slit, we have information about the radiance, Doppler shift as well as the Doppler width of the 
Ne~{\sc viii} spectral line. Time series have been obtained at this bright location by taking 
a $5\arcsec$ average over solar-Y and then wavelet power spectra have been plotted for both radiance 
and LOS velocity in Figure~\ref{fig:ne967}. 
There is a clear presence of $\approx15$~min to 20~min periodicity in 
both radiance and velocity.
Going to the off-limb inter-plume at solar-Y$\approx1020\arcsec$, time series 
in both radiance and Doppler velocity were obtained by averaging over
$9\arcsec$ in the Y direction (to increase the signal to noise ratio). 
Also from the wavelet power spectra (Figure~\ref{fig:ne1020}), a clear presence of 
$\approx15$~min to 20~min periodicity in both radiance and doppler shift was found. 
Wavelet power spectra of the 
Doppler width time series were also obtained at the two locations and are shown in Figure~\ref{fig:wdt_ne}.
The analysis reveals a periodicity similar to that observed in radiance and in Doppler shift.

EIS Fe~{\sc xii} time series have been produced by averaging over $5\arcsec$ and $9\arcsec$ in the X and 
Y-direction, respectively. Then a wavelet analysis was performed at two inter-plume locations; 
one near the limb (solar-Y$\approx1020\arcsec$) and the other further off-limb (solar-Y$\approx1120\arcsec$). 
The wavelet power spectra are shown in Figure~\ref{fig:fe1020}. 
Both heights show periodicity between 15~min and 20~min, 
consistent with the results from the x-t map (see Figure~\ref{fig:xt_fe12}). 
Furthermore, these periods are consistent with the periods
obtained from Ne~{\sc viii}. In summary, again it can be concluded that the propagating
disturbance, which originates from the on-disk bright region, propagates off-limb and in the far 
off-limb inter-plume region.

In order to check the periods of propagating disturbances in the plume region, the wavelet analysis has 
been carried out also at location solar-Y$\approx1030\arcsec$ and solar-X$\approx-39\arcsec$ by averaging 
over $9\arcsec$~in solar-Y and $5\arcsec$~in solar-X, (see Figure~\ref{fig:fe1030}). 
Also in this case, the period of propagation, $\approx$12~min to 20~min, is consistent with 
the x-t map in Figure~\ref{fig:xt_fe12p}. However, as seen from the x-t map, these disturbances
are not visible at greater heights above the limb.

\subsection{Correlation Analysis}
\begin{figure*}[htbp]
\centering
\hspace*{-0.5cm}\includegraphics[width=7.5cm]{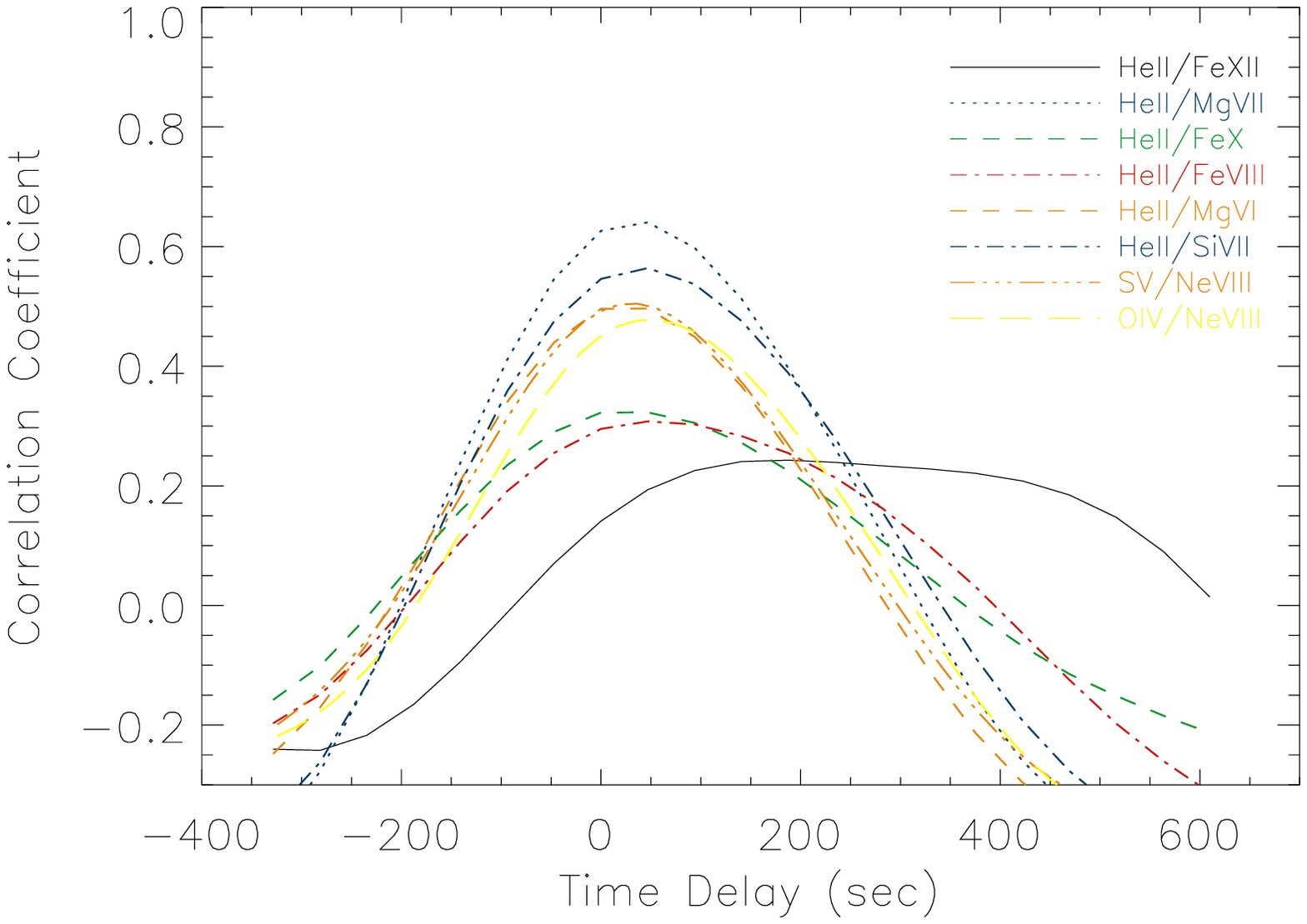}\hspace*{0.5cm}\includegraphics[width=7.5cm]{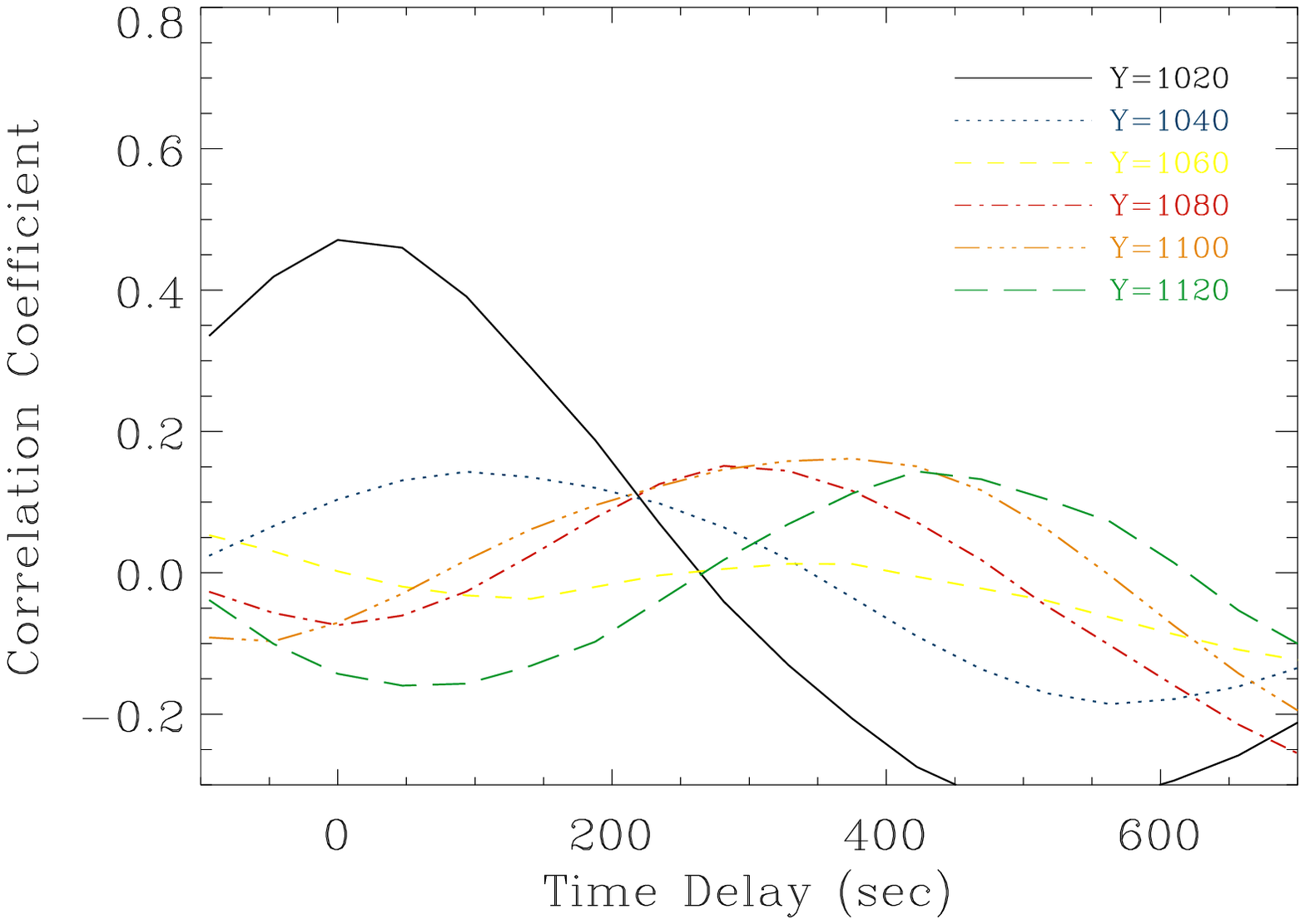}
\caption{Left: Correlation coefficients versus time delay between the line-pairs at the on-disk bright
 location (X $\approx-72\arcsec$, Y$\approx967\arcsec$). 
The maximum correlation coefficient  for a fixed line-pair provides a measure 
of the travel time. Right: Correlation coefficients versus time delay in
 inter-plume region for time series in the EIS Fe~{\sc xii} line. 
The correlation coefficients are calculated at different heights with respect to Y$\approx1000\arcsec$.}
\label{fig:correln}
\end{figure*}

In the earlier subsections we presented results from the Ne~{\sc viii} and 
Fe~{\sc xii} lines only. Propagation properties of waves can also be studied 
by correlation analysis.
First, we focus our attention on the on-disk bright region, where we 
expect the waves seen at the inter-plume location to originate. 
At this location (Y$\approx967\arcsec$), we find clear presence of 
oscillations in different lines as recorded
 by SUMER and EIS (as tabulated in Table.~\ref{tab:lines}). 
For the on-disk study, we will concentrate on the correlation between different lines as recorded by SUMER 
and EIS while for the off-limb study we will calculate correlation coefficients between different heights 
as recorded by the same Fe~{\sc xii} line. 
The correlations between the time series from two different lines
have been obtained using the IDL routine $C_{-}CORRELATE$ at different time delays between the two series. 
Correlation coefficients have been calculated for six line pairs and are plotted in the left panel of
Figure~\ref{fig:correln}. The time resolution is about 18~s for SUMER and 47~s for EIS 
(governed by the respective cadences). 
The time delay at peak of correlation can be considered as the time delay between the oscillations in the
two lines forming the pair.
It can be seen that the level of correlation and the time delay are in inverse proportion, which means
that the correlation is higher for the lines having smaller temperature separation and, because in
 coronal hole the temperature changes mainly as a function of height, smaller time delay. 
The correlation between He~{\sc ii} and Fe~{\sc xii} is smaller and shows the largest time delay whereas the
correlation between  He~{\sc ii} and Fe~{\sc x} is comparatively high and has less time delay. 
If there would be height information available, then it would
be possible to estimate the propagation speed of oscillations from one height to another. 
The radiance variation of several lines with respect to 
solar-Y has been plotted in Figure~\ref{fig:limb_var} at solar-X$\approx-72\arcsec$. As described in
 \cite{2006A&A...452.1059O}, the difference in the 
peak positions provides an estimate of the differences in the formation heights between different lines at 
that particular time and condition.
Hence, using the correlation technique, the time delay between two lines
 can be obtained and using the limb brightening technique, the formation 
height difference can be estimated for a line pair. 
The results obtained have been summarized in Table \ref{tab:correl}. 
Due to the relatively poor temporal resolution, uncertainties in time delay measurements
 are large. Hence, for the on-disk bright region,
 results obtained here can only be used to infer that waves are propagating 
from lower to higher heights in the solar atmosphere.\\

\begin{center}
\begin{table*}[htbp]
\caption{Linear correlation coefficients between oscillations in different line-pairs corresponding
 to the on-disk bright region.}
\begin{tabular}{ccccc} 
\hline \hline
Line- & log T$_{\rm max}$  & Correlation & Time delay$^{1}$ & Height diff.$^{2}$ \\
 pair & (K) & co-efficient & (s)       & (km) \\
\hline
He~{\sc ii}/Fe~{\sc xii}  & 4.9/6.1 & 0.242954 & 188 & 4509 \\
He~{\sc ii}/Fe~{\sc xi}   & 4.9/6.1 & 0.250901 & 94  & --   \\
He~{\sc ii}/Fe~{\sc x}    & 4.9/6.0 & 0.322859 & 47  & 5628 \\
He~{\sc ii}/Fe~{\sc viii} & 4.9/5.6 & 0.307703 & 47  & 3180 \\
He~{\sc ii}/Si~{\sc vii}  & 4.9/5.8 & 0.564139 & 47  & 5363 \\
He~{\sc ii}/Mg~{\sc vii}  & 4.9/5.8 & 0.640592 & 47  & 4104 \\
He~{\sc ii}/Mg~{\sc vi}   & 4.9/5.6 & 0.496354 & 47  & 3654 \\
S~{\sc v}/Ne~{\sc viii}   & 5.2/5.8 & 0.504659 & 36  & 3647 \\
O~{\sc iv}/Ne~{\sc viii}  & 5.2/5.8 & 0.477825 & 54  & 2917 \\
\hline
\end{tabular} \\
$^{1}$Limited by time resolution defind by
effective cadence of respective instruments. $^{2}$Limited by spatial resolution
of respective instruments, $\approx715$ km.
\label{tab:correl}
\end{table*}\end{center}

\begin{figure}[htbp]
\centering
\hspace*{-1cm}\includegraphics[width=7.5cm]{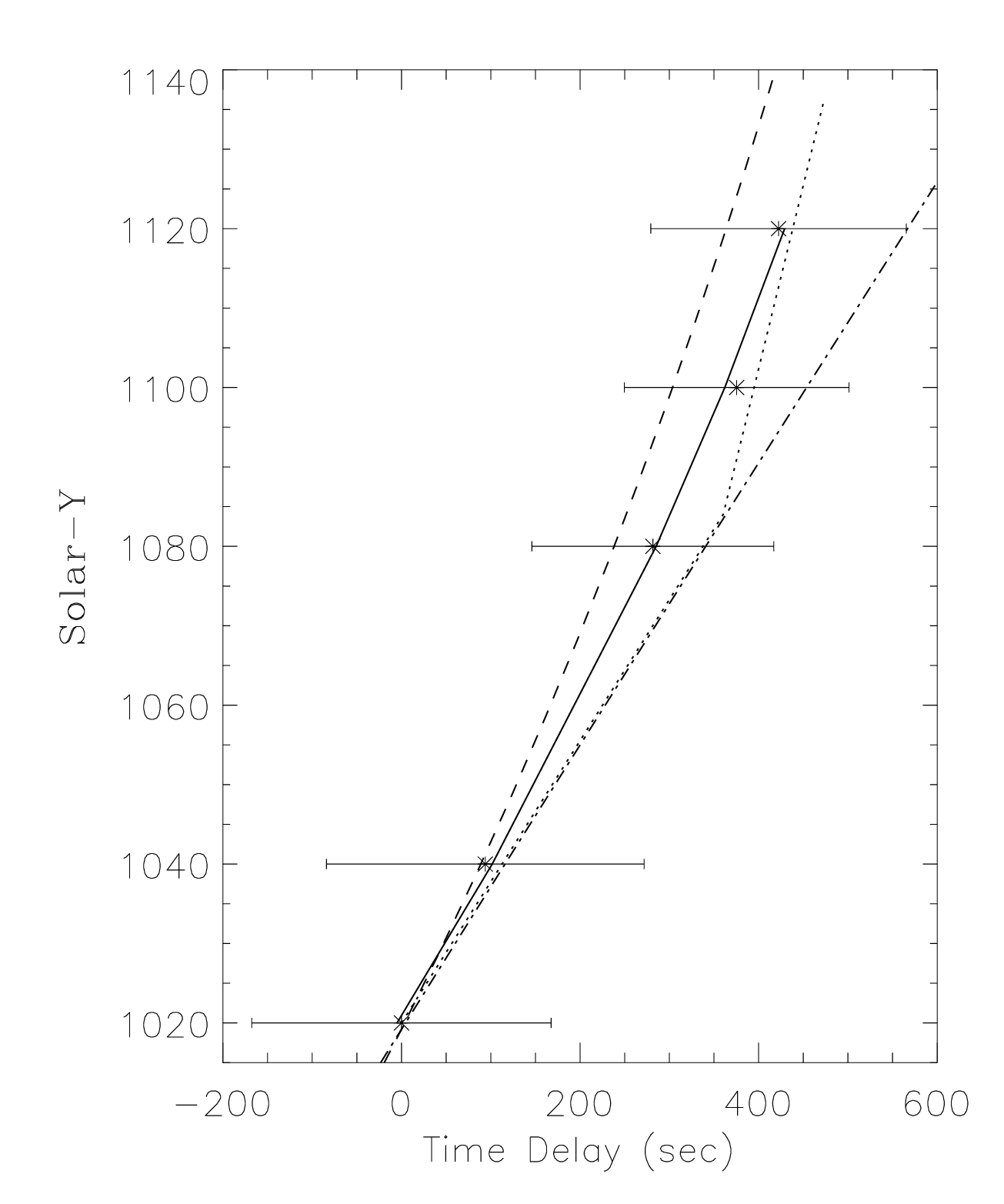}\hspace*{1cm}\includegraphics[width=6.5cm]{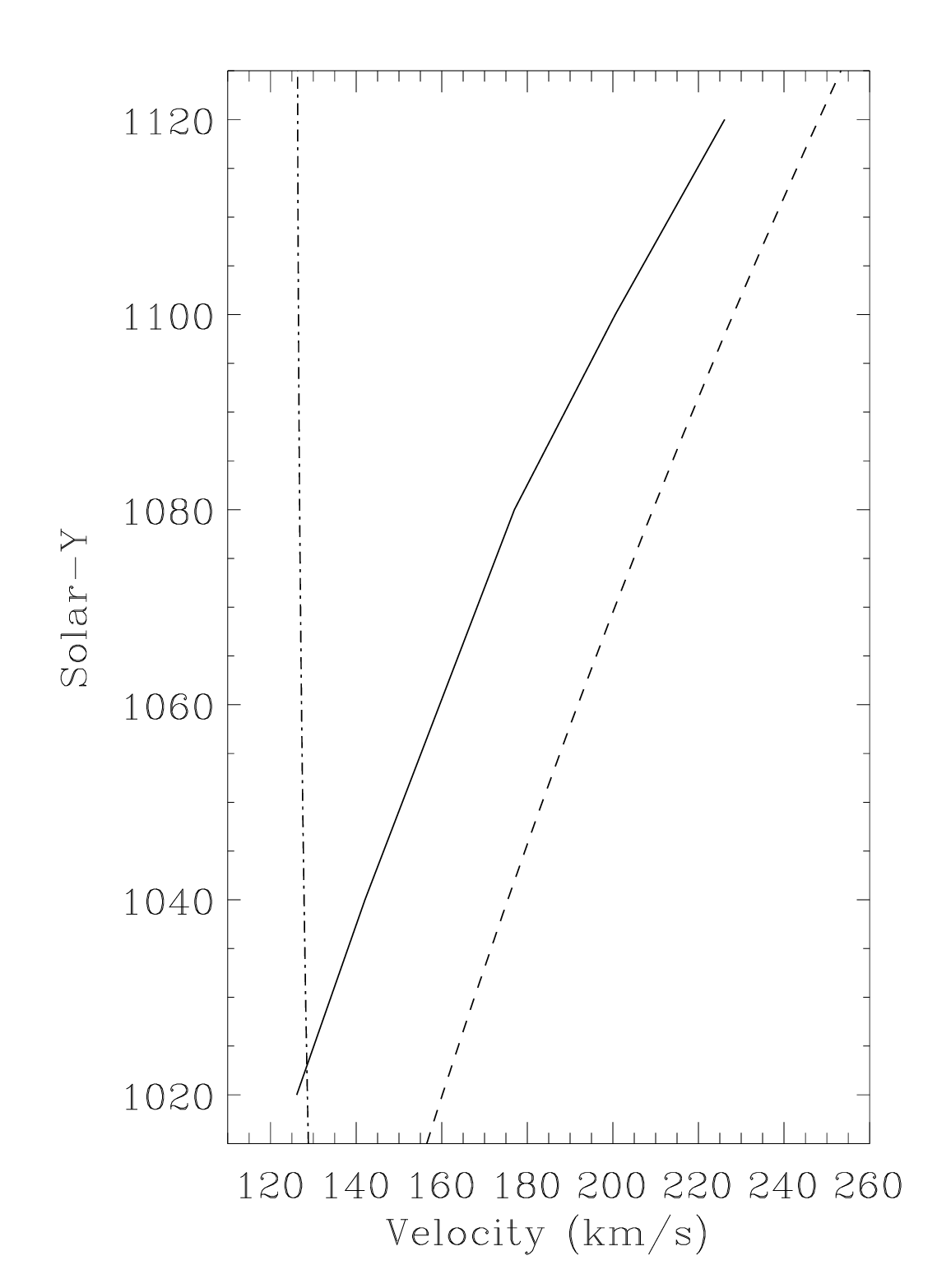}
\caption{Left: Variation of travel time with height in inter-plume region. 
The asterisks represent the measured time delays from right panel of
Figure~\ref{fig:correln} while the continuous line corresponds to a second 
order polynomial fit applied to the data points. 
The error bars on these time delays are obtained from the Half Width Half 
Maximum (HWHM) of the respective correlation peaks.
The dotted line corresponds to the fit to the slanted radiance ridges in the 
x-t maps (white lines of Figure~\ref{fig:xt_fe12}).
The dashed and the dot-dashed lines are the
theoretically predicted Alfv\'{e}nic time delays 
obtained assuming a magnetic field constant with height and expanding
according to \citet{1976SoPh...49...43K}, respectively.
The change in the slope indicates acceleration. Right: Propagation speed with height in 
inter-plume region. 
The continuous line is obtained after taking the time derivative 
of the fit to the data points in the left panel. 
The dashed and the dot-dashed lines are the
theoretically predicted Alfv\'{e}n speeds
obtained assuming a magnetic field constant with height and expanding
according to \citet{1976SoPh...49...43K}, respectively.}
\label{fig:time_speed}
\end{figure}

A similar correlation analysis is applied to the offlimb inter-plume region at
 X$\approx-72\arcsec$ using data from the same line, 
Fe~{\sc xii}, but at different heights with respect to solar-Y$\approx1000\arcsec$. 
The results are plotted in the right panel of Figure~\ref{fig:correln}. 
The time resolution is about $\approx 47$~s, the EIS cadence.
Also in this case, the level of correlations and time delays are in inverse proportion 
as expected. 
The measured time delays are plotted against solar-Y in left panel of Figure~\ref{fig:time_speed}. 
The continuous line corresponds to a second order polynomial fit applied to the data points (as marked by asterisks). 
The error bar on these time delays is obtained from the Half Width Half Maximum (HWHM) of the particular correlation plot.
The dotted line corresponds to the fit to the slanted radiance ridges in the x-t maps (white lines of 
Figure~\ref{fig:xt_fe12}). The figure indicates that the travel time is decreasing with height
indicating an acceleration, as was seen from the x-t map of Fe~{\sc xii} in Figure~\ref{fig:xt_fe12}. 
This Figure provides an independent estimate of the acceleration. 

The Alfv\'{e}n wave speed through the quasi-static corona can be 
calculated from the expression
$V_{A} = B/\sqrt{4\pi\rho}$. We use the density profile given by 
\citet{2003ApJ...588..566T} and take into account the superradial fall 
of the magnetic field with height according to
\citet{1976SoPh...49...43K}, with a base magnetic field of $\approx0.65$~G. 
For comparison purpose, the Alfv\'{e}n wave speed is
also calculated by assuming a constant magnetic field of $\approx0.65$~G with height. 
The measured time delays at different Solar-Y in inter-plume region are then compared 
with the travel time for a theoretical Alfv\'{e}n mode. 
The Alfv\'{e}nic time delays obtained assuming a magnetic field constant with height 
and expanding according to \citet{1976SoPh...49...43K}, 
are plotted, respectively, as a dashed and a dot-dashed line in the
left panel of Figure~\ref{fig:time_speed}.
To estimate the propagation speed, we plot right panel of
Figure~\ref{fig:time_speed} which shows the variation of 
propagation speed with respect to the
height in the solar atmosphere. The speed is calculated by the 
time derivative of the fit to the measured time delays 
shown in the left panel of Figure~\ref{fig:time_speed} and is plotted as a continuous line. 
This speed is compared with the
theoretically calculated propagation speed of Alfv\'{e}n modes, 
plotted as dot-dashed and dashed line, respectively, for the expanding and
constant field case. 
It can be seen that the measured
propagation speed is roughly consistent with being Alfv\'{e}nic
if we assume a field of 0.65~G at the base. 
From the figure it can be seen that
near the limb and off-limb, the speed of propagation is about 
130~km~s$^{-1}$, increasing to 
more than 220~km~s$^{-1}$ far off-limb, 
close to the speeds obtained from the x-t map in Figure~\ref{fig:xt_fe12}. 
It also appears that above a certain height (Y $\approx1080\arcsec$) 
the speed increases more rapidly. 
These results might indicate that physics of the propagation might also change
at these heights.

\section{Discussion}
\label{sec:discussion}

Our results show the propagation of disturbances from an on-disk 
region to the off-limb corona within inter-plume region. These disturbances appear 
to originate from an on-disk bright location, presumably the footpoint of a 
coronal funnel, around solar-Y$\approx967\arcsec$~(see Figure~\ref{fig:xt_ne8}).
No signature of propagation is visible below solar-Y$\approx967\arcsec$. 
The propagation speed, as measured 
from the x-t map (distance-time map) of Ne~{\sc viii} and Fe~{\sc xii} line 
radiance, increases with height. 
Outward propagating disturbances are also recorded in a plume structure 
within the EIS field of view, but the acceleration is almost zero. 
Furthermore, the disturbances become diffuse far off-limb as seen from the EIS 
Fe~{\sc xii} x-t map. 
From wavelet analysis these disturbances have a periodicity in the range of 
15~min to 20~min in both regions and the periodicity is seen almost over the whole 
duration of observation, although the power is increasing and decreasing with time. 
A correlation analysis of the light curves of  different spectral line-pairs 
in the on-disk bright region reveals larger time-delays for line pairs with the 
larger difference in formation height (inferred from the limb brightening 
curves), indicating  upward propagation. This upward propagation indicates that 
these waves are generated somewhere lower in the atmosphere, probably in the
chromosphere, since these are also seen
in He~{\sc ii} and then propagate upward towards the
off-limb region. In the off-limb region, 
cross-correlation between light curves from the same spectral line (Fe~{\sc xii}) 
but obtained at different heights above the limb shows 
time-delays indicating outward propagation, in agreement with
the results from the analysis of the x-t maps. 

The propagating disturbances, clearly visible in the inter-plume region in the 
Ne~{\sc viii} radiance x-t map, are not visible in the Ne~{\sc viii} Doppler shift and
 width x-t maps. However, the wavelet analysis
shows that both Doppler shift and width oscillations are present in the on-disk bright
 region and near off-limb with 
approximately the same periodicity as seen in radiance (see
Figures~\ref{fig:ne967}, \ref{fig:ne1020}, \& \ref{fig:wdt_ne}), and they are also mostly in 
phase. The presence of oscillations in radiance and both resolved (Doppler shift) and unresolved
 (Doppler width) velocities with approximately 
the same period, is evidence of propagating waves with at least a 
compressional component in the inter-plume lanes. 
The measured speed of propagation
of these waves from the Ne~{\sc viii} line is ($25~\pm~1.3$)~km~s$^{-1}$, 
($38~\pm~4.5$)~km~s$^{-1}$ and~($130~\pm~51$)~km~s$^{-1}$ in the 
 on-disk region, near the limb and off-limb respectively. 
Fe~{\sc xii} data shows that the propagation speed further increases to 
($330~\pm~140$)~km~s$^{-1}$ in the far off-limb region of inter-plume. 
In the plume region, instead, the observed off-limb propagation speed increases 
from ($135\pm~18$)~km~s$^{-1}$ to only ($165~\pm~43$)~km~s$^{-1}$ 
far off-limb. 
Beyond this point the radiance disturbances become diffuse. The increase in 
propagation speed is small and the acceleration is negligible within 
the given uncertainties.

The propagation speed becomes supersonic ($>C_{s}\approx170$~km~s$^{-1}$, for Fe~{\sc xii}
 line formation temperature) far off-limb in 
the inter-plume region. Moreover, near the limb region, the Ne~{\sc viii} 
and Fe~{\sc xii} lines, which are in phase, show nearly equal propagation 
speeds for some overlapping region despite having quite different formation temperatures, suggesting that 
these propagating disturbances are temperature independent.
This, together with the presence of oscillations in Doppler width and shift, may suggest that
 these waves are Alfv\'{e}nic in nature. 
Pure Alfv\'{e}n waves do not cause any density perturbation and thus would not 
cause any radiance fluctuation. However, in the real case of waves propagating 
in a density stratified atmosphere, non-linear
effect may also cause small density fluctuations leading to radiance changes of 
a few percent \citep{1995SSRv...73....1T,2009ApJ...703.1318K}. 
Moreover, oscillations in the observed line widths can be caused by torsional Alfv\'{e}n
 waves \citep{2003A&A...399L..15Z,2008ApJ...676L..73V}. Hence, interpretation of these propagating
 disturbances in terms of Alfv\'{e}n waves appears quite reasonable.

Furthermore, observed radiance oscillations can also be
explained as a line-of-sight effect of entirely incompressible 
MHD waves as described in \cite{2003A&A...397..765C}. 
In this model, when observed at an angle $\theta$ to the direction of 
propagation, the wave-induced deformation in a coronal loop causes brightness 
variations. This is because the amount of optically thin emitting plasma along the line-of-sight
 changes as a function of time. Thus, the radiance variations can be produced
even by entirely incompressible MHD waves.

Conversely, the measured propagation speeds are also consistent with
the fast magnetoacoustic mode of propagation within the error bars of the propagation speeds
 in Ne~{\sc viii} and Fe~{\sc xii} lines and explains the observed radiance oscillations due to its
compressible nature. Hence, interpretation of these propagating disturbances in
 terms of fast magnetoacoustic waves also appears reasonable.

It should be noted that it is always the apparent propagation speed in the
 plane of the sky that is measured. The structures carrying the waves most likely form an angle
 with the plane of the sky
 and, thus, these propagation speeds are always a lower limit. However, in the case of the
 inter-plume region, we trace the origin of the wave to an on-disk bright region
 located at X$\approx -72\arcsec$ and Y$\approx 967\arcsec$. Assuming roughly radial propagation, this would
 imply an angle of $\approx 9\degree$ with the plane of sky, leading to a difference between
 apparent and real propagation speed of less than $2~\%$, a negligible quantity.  In the case of
 the plume, a careful inspection of Figure~\ref{fig:stereo} shows that the plume is
 rooted very close to the visible limb and is lying
 close to the plane of the sky. This suggests that in this case the difference between apparent and 
real propagation speed is likely very small.

The kind of oscillations which are presented here are very similar to other 
already reported in the literature \citep[e.g.][]{1998ApJ...501L.217D,
2000ApJ...529..592O,2000SoPh..196...63B,2001A&A...380L..39B,2004ApJ...605..521M,
2006A&A...452.1059O,2007A&A...463..713O,2009A&A...493..251G,2009A&A...499L..29B}
in on-disk and off-limb regions of coronal holes observed with different 
instruments. \citet{2001A&A...380L..39B} have reported oscillations in 
on-disk network regions and in inter-plume regions similar to those reported here. 
\citet{2009A&A...493..251G} have calculated the speed of propagation in the 
on-disk bright region using the statistical technique and the measured speed 
were of the same order as reported here. 
All these observations indicate the presence of propagating MHD waves. 
Recently, \citet{2010A&A...510L...2M} detected propagating 
features in polar plumes using STEREO observations. 
These authors have interpreted these features in terms of high speed jets of 
plasma traveling along the structures which repeat quasi-periodically, with 
repeat-times ranging from 5~min to 25~min. This would be contrary to the widely 
held interpretation that this observational phenomenon is due to compressive waves. 
One should also consider Doppler dimming results from UVCS and SUMER aboard 
SoHO, which shows that velocities above 100~km~s$^{-1}$ are reached only above 
1.5~R/R$_{\odot}$ in either plumes or inter-plumes 
\citep{2003ApJ...588..566T,2003ApJ...589..623G,2005ApJ...635L.185G}.\\

To our knowledge, this is the first 
time that a signature of accelerating Alfv\'{e}nic waves or fast magnetoacoustic waves 
originating in an on-disk bright region has been observed
in near and far off-limb regions within 1.2~R/R$_{\odot}$.
In inter-plume region, the disturbance
is seen propagating up along the whole EIS slot length whereas in the plume region, 
it becomes diffuse far off-limb.
In inter-plume region the wave propagates farther in the 
corona with high acceleration whereas in plume region it may have been dissipated in 
the off-limb region. 
However, the lack of signature at greater heights in the
plume may be simply due to merging with the background signal.
This suggests that inter-plume regions may be the preferred  channel for the 
acceleration of the fast solar wind. 
This conclusion is in agreement with earlier reports 
\citep{1997ApJ...484L..75W,2000A&A...359L...1P,2000ApJ...531L..79G,
2001A&A...380L..39B,2003ApJ...588..566T}.


\section{Conclusion}
\label{sec:conclusion}

Analysis of Ne~{\sc viii} and Fe~{\sc xii} radiance x-t maps reveal the 
presence of outward propagating radiance disturbances in the off-limb and near 
off-limb region of inter-plume with periodicities of about 15~min to 20~min.
From the SUMER Ne~{\sc viii} line radiance x-t map, one can infer that the 
waves originate from a bright location (presumably the footpoint of a coronal 
funnel) and propagate towards the limb with a speed 
($25~\pm~1.3$)~km~s$^{-1}$. 
Around the limb the speed has increased to ($38~\pm~4.5$)~km~s$^{-1}$,
reaching ($130~\pm~51$)~km~s$^{-1}$ off-limb.
Further far off-limb, the speed of the propagation
becomes ($330~\pm~140$)~km~s$^{-1}$ as seen in the EIS Fe~{\sc xii} line. 
Similar propagating disturbances are also seen in the plume region but with 
negligible acceleration, if any. 
The waves are not
visible far off-limb, suggesting that they may be dissipated or, more simply,
merge into the background.
The waves as recorded in the inter-plume regions are either Alfv\'{e}nic or
 fast magnetoacoustic in nature 
whereas the one seen in plumes are more likely slow magnetoacoustic type. 
\citet{2005Sci...308..519T} have conjectured that 
the solar wind outflow is launched by reconnection at network boundaries 
between open flux lines and intra-network closed loops. The intra-network 
closed loops are pushed by supergranular convection towards the network 
triggering reconnection. This scenario is consistent with our identification 
of the origin of the propagating disturbances in inter-plume region with an on-disk 
bright region. 
These results support the view that the inter-plume regions are the preferred 
channel for the acceleration of the fast solar wind.   


\begin{acknowledgements} 
We are grateful to the anonymous referee for valuable comments and suggestions which
 improved the quality of the presentation. This work was supported by the Indo-German DST-DAAD
 joint project D/07/03045. 
The SUMER project is financially supported by DLR, CNES, NASA, and the ESA  
PRODEX programme (Swiss contribution). 
Hinode is a Japanese mission developed and launched by ISAS/JAXA, with NAOJ 
as domestic partner and NASA and STFC (UK) as international partners.  
It is operated by these agencies in co-operation with ESA and NSC (Norway).
This work was partially supported by the WCU grant No. R31-10016 from the 
Korean Ministry of Education, Science and Technology.
\end{acknowledgements}

\bibliographystyle{apj.bst}
\bibliography{references}

\begin{thebibliography}{62}
\expandafter\ifx\csname natexlab\endcsname\relax\def\natexlab#1{#1}\fi

\bibitem[{{Ahmad} \& {Withbroe}(1977)}]{1977SoPh...53..397A}
{Ahmad}, I.~A., \& {Withbroe}, G.~L. 1977, \solphys, 53, 397

\bibitem[{{Antonucci} {et~al.}(2000){Antonucci}, {Dodero}, \&
  {Giordano}}]{2000SoPh..197..115A}
{Antonucci}, E., {Dodero}, M.~A., \& {Giordano}, S. 2000, \solphys, 197, 115

\bibitem[{{Antonucci} {et~al.}(2004){Antonucci}, {Dodero}, {Giordano},
  {Krishnakumar}, \& {Noci}}]{2004A&A...416..749A}
{Antonucci}, E., {Dodero}, M.~A., {Giordano}, S., {Krishnakumar}, V., \&
  {Noci}, G. 2004, \aap, 416, 749

\bibitem[{{Banerjee} {et~al.}(2000{\natexlab{a}}){Banerjee}, {O'Shea}, \&
  {Doyle}}]{2000SoPh..196...63B}
{Banerjee}, D., {O'Shea}, E., \& {Doyle}, J.~G. 2000{\natexlab{a}}, \solphys,
  196, 63

\bibitem[{{Banerjee} {et~al.}(2001){Banerjee}, {O'Shea}, {Doyle}, \&
  {Goossens}}]{2001A&A...380L..39B}
{Banerjee}, D., {O'Shea}, E., {Doyle}, J.~G., \& {Goossens}, M. 2001, \aap,
  380, L39

\bibitem[{{Banerjee} {et~al.}(2009{\natexlab{a}}){Banerjee},
  {P{\'e}rez-Su{\'a}rez}, \& {Doyle}}]{2009A&A...501L..15B}
{Banerjee}, D., {P{\'e}rez-Su{\'a}rez}, D., \& {Doyle}, J.~G.
  2009{\natexlab{a}}, \aap, 501, L15

\bibitem[{{Banerjee} {et~al.}(2000{\natexlab{b}}){Banerjee}, {Teriaca},
  {Doyle}, \& {Lemaire}}]{2000SoPh..194...43B}
{Banerjee}, D., {Teriaca}, L., {Doyle}, J.~G., \& {Lemaire}, P.
  2000{\natexlab{b}}, \solphys, 194, 43

\bibitem[{{Banerjee} {et~al.}(1998){Banerjee}, {Teriaca}, {Doyle}, \&
  {Wilhelm}}]{1998A&A...339..208B}
{Banerjee}, D., {Teriaca}, L., {Doyle}, J.~G., \& {Wilhelm}, K. 1998, \aap,
  339, 208

\bibitem[{{Banerjee} {et~al.}(2009{\natexlab{b}}){Banerjee}, {Teriaca},
  {Gupta}, {Imada}, {Stenborg}, \& {Solanki}}]{2009A&A...499L..29B}
{Banerjee}, D., {Teriaca}, L., {Gupta}, G.~R., {Imada}, S., {Stenborg}, G., \&
  {Solanki}, S.~K. 2009{\natexlab{b}}, \aap, 499, L29

\bibitem[{{Belcher}(1971)}]{1971ApJ...168..509B}
{Belcher}, J.~W. 1971, \apj, 168, 509

\bibitem[{{Bohlin} {et~al.}(1975){Bohlin}, {Sheeley}, \&
  {Tousey}}]{1975spre.conf..651B}
{Bohlin}, J.~D., {Sheeley}, N.~R., \& {Tousey}, R. 1975, in Space Research XV,
  ed. {M.~J.~Rycroft}, 651--656

\bibitem[{{Casalbuoni} {et~al.}(1999){Casalbuoni}, {Del Zanna}, {Habbal}, \&
  {Velli}}]{1999JGR...104.9947C}
{Casalbuoni}, S., {Del Zanna}, L., {Habbal}, S.~R., \& {Velli}, M. 1999, \jgr,
  104, 9947

\bibitem[{{Cooper} {et~al.}(2003){Cooper}, {Nakariakov}, \&
  {Tsiklauri}}]{2003A&A...397..765C}
{Cooper}, F.~C., {Nakariakov}, V.~M., \& {Tsiklauri}, D. 2003, \aap, 397, 765

\bibitem[{{Cranmer}(2009)}]{2009LRSP....6....3C}
{Cranmer}, S.~R. 2009, Living Reviews in Solar Physics, 6, 3

\bibitem[{{Culhane} {et~al.}(2007){Culhane}, {Harra}, {James}, {Al-Janabi},
  {Bradley}, {Chaudry}, {Rees}, {Tandy}, {Thomas}, {Whillock}, {Winter},
  {Doschek}, {Korendyke}, {Brown}, {Myers}, {Mariska}, {Seely}, {Lang}, {Kent},
  {Shaughnessy}, {Young}, {Simnett}, {Castelli}, {Mahmoud}, {Mapson-Menard},
  {Probyn}, {Thomas}, {Davila}, {Dere}, {Windt}, {Shea}, {Hagood}, {Moye},
  {Hara}, {Watanabe}, {Matsuzaki}, {Kosugi}, {Hansteen}, \&
  {Wikstol}}]{2007SoPh..243...19C}
{Culhane}, J.~L., {et~al.} 2007, \solphys, 243, 19

\bibitem[{{De Pontieu} {et~al.}(2007{\natexlab{a}}){De Pontieu}, {McIntosh},
  {Hansteen}, {Carlsson}, {Schrijver}, {Tarbell}, {Title}, {Shine}, {Suematsu},
  {Tsuneta}, {Katsukawa}, {Ichimoto}, {Shimizu}, \&
  {Nagata}}]{2007PASJ...59S.655D}
{De Pontieu}, B., {et~al.} 2007{\natexlab{a}}, \pasj, 59, 655

\bibitem[{{De Pontieu} {et~al.}(2007{\natexlab{b}}){De Pontieu}, {McIntosh},
  {Carlsson}, {Hansteen}, {Tarbell}, {Schrijver}, {Title}, {Shine}, {Tsuneta},
  {Katsukawa}, {Ichimoto}, {Suematsu}, {Shimizu}, \&
  {Nagata}}]{2007Sci...318.1574D}
---. 2007{\natexlab{b}}, Science, 318, 1574

\bibitem[{{DeForest} \& {Gurman}(1998)}]{1998ApJ...501L.217D}
{DeForest}, C.~E., \& {Gurman}, J.~B. 1998, \apjl, 501, L217+

\bibitem[{{DeForest} {et~al.}(1997){DeForest}, {Hoeksema}, {Gurman},
  {Thompson}, {Plunkett}, {Howard}, {Harrison}, \&
  {Hassler}}]{1997SoPh..175..393D}
{DeForest}, C.~E., {Hoeksema}, J.~T., {Gurman}, J.~B., {Thompson}, B.~J.,
  {Plunkett}, S.~P., {Howard}, R., {Harrison}, R.~C., \& {Hassler}, D.~M. 1997,
  \solphys, 175, 393

\bibitem[{{Dolla} \& {Solomon}(2008)}]{2008A&A...483..271D}
{Dolla}, L., \& {Solomon}, J. 2008, \aap, 483, 271

\bibitem[{{Gabriel} {et~al.}(2005){Gabriel}, {Abbo}, {Bely-Dubau}, {Llebaria},
  \& {Antonucci}}]{2005ApJ...635L.185G}
{Gabriel}, A.~H., {Abbo}, L., {Bely-Dubau}, F., {Llebaria}, A., \& {Antonucci},
  E. 2005, \apjl, 635, L185

\bibitem[{{Gabriel} {et~al.}(2003){Gabriel}, {Bely-Dubau}, \&
  {Lemaire}}]{2003ApJ...589..623G}
{Gabriel}, A.~H., {Bely-Dubau}, F., \& {Lemaire}, P. 2003, \apj, 589, 623

\bibitem[{{Giordano} {et~al.}(2000){Giordano}, {Antonucci}, {Noci}, {Romoli},
  \& {Kohl}}]{2000ApJ...531L..79G}
{Giordano}, S., {Antonucci}, E., {Noci}, G., {Romoli}, M., \& {Kohl}, J.~L.
  2000, \apjl, 531, L79

\bibitem[{{Gupta} {et~al.}(2009){Gupta}, {O'Shea}, {Banerjee}, {Popescu}, \&
  {Doyle}}]{2009A&A...493..251G}
{Gupta}, G.~R., {O'Shea}, E., {Banerjee}, D., {Popescu}, M., \& {Doyle}, J.~G.
  2009, \aap, 493, 251

\bibitem[{{Howard} {et~al.}(2008){Howard}, {Moses}, {Vourlidas}, {Newmark},
  {Socker}, {Plunkett}, {Korendyke}, {Cook}, {Hurley}, {Davila}, {Thompson},
  {St Cyr}, {Mentzell}, {Mehalick}, {Lemen}, {Wuelser}, {Duncan}, {Tarbell},
  {Wolfson}, {Moore}, {Harrison}, {Waltham}, {Lang}, {Davis}, {Eyles},
  {Mapson-Menard}, {Simnett}, {Halain}, {Defise}, {Mazy}, {Rochus}, {Mercier},
  {Ravet}, {Delmotte}, {Auchere}, {Delaboudiniere}, {Bothmer}, {Deutsch},
  {Wang}, {Rich}, {Cooper}, {Stephens}, {Maahs}, {Baugh}, {McMullin}, \&
  {Carter}}]{2008SSRv..136...67H}
{Howard}, R.~A., {et~al.} 2008, Space Science Reviews, 136, 67

\bibitem[{{Imada}(2010)}]{2010Imada}
{Imada}, S. 2010, in preparation

\bibitem[{{Jess} {et~al.}(2009){Jess}, {Mathioudakis}, {Erd{\'e}lyi},
  {Crockett}, {Keenan}, \& {Christian}}]{2009Sci...323.1582J}
{Jess}, D.~B., {Mathioudakis}, M., {Erd{\'e}lyi}, R., {Crockett}, P.~J.,
  {Keenan}, F.~P., \& {Christian}, D.~J. 2009, Science, 323, 1582

\bibitem[{{Kaghashvili} {et~al.}(2009){Kaghashvili}, {Quinn}, \&
  {Hollweg}}]{2009ApJ...703.1318K}
{Kaghashvili}, E.~K., {Quinn}, R.~A., \& {Hollweg}, J.~V. 2009, \apj, 703, 1318

\bibitem[{{Kopp} \& {Holzer}(1976)}]{1976SoPh...49...43K}
{Kopp}, R.~A., \& {Holzer}, T.~E. 1976, \solphys, 49, 43

\bibitem[{{Kosugi} {et~al.}(2007){Kosugi}, {Matsuzaki}, {Sakao}, {Shimizu},
  {Sone}, {Tachikawa}, {Hashimoto}, {Minesugi}, {Ohnishi}, {Yamada}, {Tsuneta},
  {Hara}, {Ichimoto}, {Suematsu}, {Shimojo}, {Watanabe}, {Shimada}, {Davis},
  {Hill}, {Owens}, {Title}, {Culhane}, {Harra}, {Doschek}, \&
  {Golub}}]{2007SoPh..243....3K}
{Kosugi}, T., {et~al.} 2007, \solphys, 243, 3

\bibitem[{{Krieger} {et~al.}(1973){Krieger}, {Timothy}, \&
  {Roelof}}]{1973SoPh...29..505K}
{Krieger}, A.~S., {Timothy}, A.~F., \& {Roelof}, E.~C. 1973, \solphys, 29, 505

\bibitem[{{Landi} \& {Cranmer}(2009)}]{2009ApJ...691..794L}
{Landi}, E., \& {Cranmer}, S.~R. 2009, \apj, 691, 794

\bibitem[{{McComas} {et~al.}(2000){McComas}, {Barraclough}, {Funsten},
  {Gosling}, {Santiago-Mu{\~n}oz}, {Skoug}, {Goldstein}, {Neugebauer}, {Riley},
  \& {Balogh}}]{2000JGR...10510419M}
{McComas}, D.~J., {et~al.} 2000, \jgr, 105, 10419

\bibitem[{{McIntosh} {et~al.}(2010){McIntosh}, {Innes}, {de Pontieu}, \&
  {Leamon}}]{2010A&A...510L...2M}
{McIntosh}, S.~W., {Innes}, D.~E., {de Pontieu}, B., \& {Leamon}, R.~J. 2010,
  \aap, 510, L2+

\bibitem[{{Morgan} {et~al.}(2004){Morgan}, {Habbal}, \&
  {Li}}]{2004ApJ...605..521M}
{Morgan}, H., {Habbal}, S.~R., \& {Li}, X. 2004, \apj, 605, 521

\bibitem[{{Munro} \& {Withbroe}(1972)}]{1972ApJ...176..511M}
{Munro}, R.~H., \& {Withbroe}, G.~L. 1972, \apj, 176, 511

\bibitem[{{Noci} {et~al.}(1997){Noci}, {Kohl}, {Antonucci}, {Tondello},
  {Huber}, {Fineschi}, {Gardner}, {Naletto}, {Nicolosi}, {Raymond}, {Romoli},
  {Spadaro}, {Siegmund}, {Benna}, {Ciaravella}, {Giordano}, {Michels},
  {Modigliani}, {Panasyuk}, {Pernechele}, {Poletto}, {Smith}, \&
  {Strachan}}]{1997AdSpR..20.2219N}
{Noci}, G., {et~al.} 1997, Advances in Space Research, 20, 2219

\bibitem[{{Ofman}(2005)}]{2005SSRv..120...67O}
{Ofman}, L. 2005, Space Science Reviews, 120, 67

\bibitem[{{Ofman} {et~al.}(1997){Ofman}, {Romoli}, {Poletto}, {Noci}, \&
  {Kohl}}]{1997ApJ...491L.111O}
{Ofman}, L., {Romoli}, M., {Poletto}, G., {Noci}, G., \& {Kohl}, J.~L. 1997,
  \apjl, 491, L111+

\bibitem[{{Ofman} {et~al.}(2000){Ofman}, {Romoli}, {Poletto}, {Noci}, \&
  {Kohl}}]{2000ApJ...529..592O}
---. 2000, \apj, 529, 592

\bibitem[{{O'Shea} {et~al.}(2006){O'Shea}, {Banerjee}, \&
  {Doyle}}]{2006A&A...452.1059O}
{O'Shea}, E., {Banerjee}, D., \& {Doyle}, J.~G. 2006, \aap, 452, 1059

\bibitem[{{O'Shea} {et~al.}(2007){O'Shea}, {Banerjee}, \&
  {Doyle}}]{2007A&A...463..713O}
---. 2007, \aap, 463, 713

\bibitem[{{Patsourakos} \& {Vial}(2000)}]{2000A&A...359L...1P}
{Patsourakos}, S., \& {Vial}, J.-C. 2000, \aap, 359, L1

\bibitem[{{Popescu} {et~al.}(2005){Popescu}, {Banerjee}, {O'Shea}, {Doyle}, \&
  {Xia}}]{2005A&A...442.1087P}
{Popescu}, M.~D., {Banerjee}, D., {O'Shea}, E., {Doyle}, J.~G., \& {Xia}, L.~D.
  2005, \aap, 442, 1087

\bibitem[{{Raouafi} {et~al.}(2007){Raouafi}, {Harvey}, \&
  {Solanki}}]{2007ApJ...658..643R}
{Raouafi}, N., {Harvey}, J.~W., \& {Solanki}, S.~K. 2007, \apj, 658, 643

\bibitem[{{Suzuki} \& {Inutsuka}(2005)}]{2005ApJ...632L..49S}
{Suzuki}, T.~K., \& {Inutsuka}, S. 2005, \apjl, 632, L49

\bibitem[{{Telloni} {et~al.}(2007){Telloni}, {Antonucci}, \&
  {Dodero}}]{2007A&A...472..299T}
{Telloni}, D., {Antonucci}, E., \& {Dodero}, M.~A. 2007, \aap, 472, 299

\bibitem[{{Teriaca} {et~al.}(1999){Teriaca}, {Banerjee}, \&
  {Doyle}}]{1999A&A...349..636T}
{Teriaca}, L., {Banerjee}, D., \& {Doyle}, J.~G. 1999, \aap, 349, 636

\bibitem[{{Teriaca} {et~al.}(2003){Teriaca}, {Poletto}, {Romoli}, \&
  {Biesecker}}]{2003ApJ...588..566T}
{Teriaca}, L., {Poletto}, G., {Romoli}, M., \& {Biesecker}, D.~A. 2003, \apj,
  588, 566

\bibitem[{{Tomczyk} {et~al.}(2007){Tomczyk}, {McIntosh}, {Keil}, {Judge},
  {Schad}, {Seeley}, \& {Edmondson}}]{2007Sci...317.1192T}
{Tomczyk}, S., {McIntosh}, S.~W., {Keil}, S.~L., {Judge}, P.~G., {Schad}, T.,
  {Seeley}, D.~H., \& {Edmondson}, J. 2007, Science, 317, 1192

\bibitem[{{Torrence} \& {Compo}(1998)}]{1998BAMS...79...61T}
{Torrence}, C., \& {Compo}, G.~P. 1998, Bulletin of the American Meteorological
  Society, 79, 61

\bibitem[{{Tu} \& {Marsch}(1995)}]{1995SSRv...73....1T}
{Tu}, C., \& {Marsch}, E. 1995, Space Science Reviews, 73, 1

\bibitem[{{Tu} {et~al.}(2005){Tu}, {Zhou}, {Marsch}, {Xia}, {Zhao}, {Wang}, \&
  {Wilhelm}}]{2005Sci...308..519T}
{Tu}, C.-Y., {Zhou}, C., {Marsch}, E., {Xia}, L.-D., {Zhao}, L., {Wang}, J.-X.,
  \& {Wilhelm}, K. 2005, Science, 308, 519

\bibitem[{{Van Doorsselaere} {et~al.}(2008){Van Doorsselaere}, {Nakariakov}, \&
  {Verwichte}}]{2008ApJ...676L..73V}
{Van Doorsselaere}, T., {Nakariakov}, V.~M., \& {Verwichte}, E. 2008, \apjl,
  676, L73

\bibitem[{{Wang} {et~al.}(1997){Wang}, {Sheeley}, {Dere}, {Duffin}, {Howard},
  {Michels}, {Moses}, {Harvey}, {Branston}, {Delaboudiniere}, {Artzner},
  {Hochedez}, {Defise}, {Catura}, {Lemen}, {Gurman}, {Neupert}, {Newmark},
  {Thompson}, \& {Maucherat}}]{1997ApJ...484L..75W}
{Wang}, Y.-M., {et~al.} 1997, \apjl, 484, L75+

\bibitem[{{Wilhelm}(2006)}]{2006A&A...455..697W}
{Wilhelm}, K. 2006, \aap, 455, 697

\bibitem[{{Wilhelm} {et~al.}(2000){Wilhelm}, {Dammasch}, {Marsch}, \&
  {Hassler}}]{2000A&A...353..749W}
{Wilhelm}, K., {Dammasch}, I.~E., {Marsch}, E., \& {Hassler}, D.~M. 2000, \aap,
  353, 749

\bibitem[{{Wilhelm} {et~al.}(1995){Wilhelm}, {Curdt}, {Marsch}, {Sch{\"u}hle},
  {Lemaire}, {Gabriel}, {Vial}, {Grewing}, {Huber}, {Jordan}, {Poland},
  {Thomas}, {K{\"u}hne}, {Timothy}, {Hassler}, \&
  {Siegmund}}]{1995SoPh..162..189W}
{Wilhelm}, K., {et~al.} 1995, \solphys, 162, 189

\bibitem[{{Wilhelm} {et~al.}(1997){Wilhelm}, {Lemaire}, {Curdt}, {Schuhle},
  {Marsch}, {Poland}, {Jordan}, {Thomas}, {Hassler}, {Huber}, {Vial}, {Kuhne},
  {Siegmund}, {Gabriel}, {Timothy}, {Grewing}, {Feldman}, {Hollandt}, \&
  {Brekke}}]{1997SoPh..170...75W}
---. 1997, \solphys, 170, 75

\bibitem[{{Woch} {et~al.}(1997){Woch}, {Axford}, {Mall}, {Wilken}, {Livi},
  {Geiss}, {Gloeckler}, \& {Forsyth}}]{1997GeoRL..24.2885W}
{Woch}, J., {Axford}, W.~I., {Mall}, U., {Wilken}, B., {Livi}, S., {Geiss}, J.,
  {Gloeckler}, G., \& {Forsyth}, R.~J. 1997, \grl, 24, 2885

\bibitem[{{Zaqarashvili}(2003)}]{2003A&A...399L..15Z}
{Zaqarashvili}, T.~V. 2003, \aap, 399, L15

\bibitem[{{Zhang} {et~al.}(2003){Zhang}, {Woch}, {Solanki}, {von Steiger}, \&
  {Forsyth}}]{2003JGRA..108.1144Z}
{Zhang}, J., {Woch}, J., {Solanki}, S.~K., {von Steiger}, R., \& {Forsyth}, R.
  2003, Journal of Geophysical Research (Space Physics), 108, 1144

\end{thebibliography}

\end{document}